\documentclass[british]{article}
\usepackage[T1]{fontenc}
\usepackage[latin9]{inputenc}
\usepackage{geometry}
\usepackage{units}
\usepackage{amsmath}
\usepackage{amsthm}
\usepackage{amssymb}
\usepackage{setspace}
\usepackage{graphicx}
\usepackage{enumitem}
\usepackage{booktabs}
\usepackage{xcolor}

\makeatletter

\providecommand{\tabularnewline}{\\}

\theoremstyle{plain}
    \ifx\thechapter\undefined
      \newtheorem{assumption}{\protect\assumptionname}
    \else
      \newtheorem{assumption}{\protect\assumptionname}[chapter]
    \fi

\newtheorem{proposition}{Proposition}

\@ifundefined{showcaptionsetup}{}{%
 \PassOptionsToPackage{caption=false}{subfig}}
\usepackage{subfig}
\makeatother

\usepackage{babel}

\newcommand{\eqTwo}[2]{\stackrel{(\ref{#1}),(\ref{#2})}{=}}

\providecommand{\assumptionname}{Assumption}

\newtheorem{remark}{Remark}

\title{Tracking control for a cascade perturbed control system using active disturbance rejection paradigm\thanks{This work was supported by the National Science Centre (NCN) under the grant No. 2014/15/B/ST7/00429, contract No. UMO-2014/15/B/ST7/00429.}}

\author{Rados{\l}aw Patelski, Dariusz Pazderski\\
  Pozna\'n University of Technology \\
  Institute of Automation and Robotics \\
  ul. Piotrowo 3a~~60-965 Pozna\'n, Poland
}

\begin{document}
\maketitle
\begin{abstract}
In this paper the stability of a closed-loop cascade control system in the trajectory tracking task is addressed. The considered plant consists of underlying second-order fully actuated perturbed dynamics and the first order system which describes dynamics of the input. The main theoretical result presented in the paper concerns stability conditions formulated based on the Lyapunov analysis for the cascade control structure taking advantage of the active rejection disturbance approach. In particular, limitations imposed on a feasible set of an observer bandwidth are discussed. In order to illustrate characteristics of the closed-loop control system simulation results are presented. Furthermore, the controller is verified experimentally using a two-axis telescope mount. The obtained results confirm that the considered control strategy can be efficiently applied for mechanical systems when a high tracking precision is required. 
\end{abstract}

\section{Introduction}
Set-point regulation and trajectory tracking  constitute elementary tasks in control theory. It is well known that a fundamental method of stabilisation by means of a smooth static state feedback has significant limitations which come, among others, from the inability to measure the state as well as the occurrence of parametric and structural model uncertainties. Thus, for these reasons, various adaptive and robust control techniques are required to improve the performance of the closed-loop system. In particular, algorithms used for the state and disturbance estimation are of great importance here.

The use of high gain observers (HGOs) is well motivated in the theory of linear dynamic systems, where it is commonly assumed that state estimation dynamics are  negligible with respect to the dominant dynamics of the closed-loop system. A similar approach can be employed successfully for a certain class of nonlinear systems where establishing a fast convergence of estimation errors may be sufficient to ensure the stability, \cite{KhP:2014}. In a natural way, the HGO observer is a basic tool to support a control feedback when a plant model is roughly known. Here one can mention the free-model control paradigm introduced by Fliess and others, \cite{Fliess:2009,FlJ:2013} as well as the active disturbance rejection control (ADRC) proposed by Han and Gao, \cite{Han:1998,Gao:2002,Gao:2006,Han:2009}. 

It turns out that the above-mentioned  control methodology can be highly competitive with respect to the classic PID technique in many industrial applications, \cite{SiGao:2005,WCW:2007,MiGao:2005,CZG:2007,MiH:2015,NSKCFL:2018}. 
Furthermore, it can be regarded as an alternative control approach in comparison to the sliding control technique proposed by Utkin and others, \cite{Utk:77, Bartol:2008}, where bounded matched disturbances are rejected due to fast switching discontinuous controls. Thus, it is possible to stabilise the closed-loop control system, in the sense of Filippov, on a prescribed, possibly time-varying, sliding surface, \cite{Bart:96, NVMPB:2012}. Currently, also second and higher-order sliding techniques for control and state estimations are being explored, \cite{Levant:1993, Levant:1998, Bartol:1998, Cast:2016}. It is  noteworthy to recall a recent control algorithm based on higher-order sliding modes to solve the tracking problem in a finite time for a class of uncertain mechanical systems in robotics, \cite{Gal:2015, Gal:2016}. From a theoretical point of view, some questions arise regarding conditions of application of control techniques based on a disturbance observer, with particular emphasis on maintaining the stability of the closed-loop system. Recently, new results concerning this issue have been reported for ADRC controllers, \cite{SiGao:2017,ACSA:2017}. In this paper we further study the ADRC methodology taking into account a particular structure of perturbed plant. Basically, we deal with a cascade control system which is composed of two parts. The first component is represented by  second-order dynamics which constitute an essential part of the plant. It is assumed that the system is fully actuated and subject to matched-type disturbances with bounded partial derivatives. The second component is defined by an elementary first-order linear system which describes input dynamics of the entire plant. Simultaneously, it is supposed that the state and control input of the second order dynamics are not fully available. 

It can be seen that the considered plant well corresponds to a class of mechanical systems equipped with a local feedback applied at the level of actuators. As a result of additional dynamics, real control forces are not accessible directly which may deteriorate the stability of the closed-loop system.

In order to analyse the closed-loop system we take advantage of Lyapunov tools. Basically, we investigate how an extended state observer (ESO) affects the stability when additional input dynamics are considered. Further we formulate stability conditions and estimate bounds of errors. In particular, we show that the observer gains cannot be made arbitrarily large as it is commonly recommended in the ADRC paradigm. Such an obstruction is a result of the occurrence of input dynamics which is not explicitly taken into account in the feedback design procedure. 

According to the best authors' knowledge, the Lyapunov stability analysis for the considered control structure taking advantage of the ADRC approach has not been addressed in the literature so far. 

Theoretical results are illustrated by numerical simulations and experiments. The experimental validation are conducted on a real two-axis telescope mount driven by synchronous gearless motors, \cite{KPKJPKBJN:2019}. Here we show that the considered methods provide high tracking accuracy which is required in such an application. Additionally, we compare the efficiency of compensation terms, computed based on the reference trajectory and on-line estimates in order to improve the tracking performance.

The paper is organised as follows. In Section 2 the model of a cascade control process is introduced. Then a preliminary feedback is designed and a corresponding extended state observer is proposed. The stability of the closed-loop system is studied using Lyapunov tools and stability conditions with respect to the considered control structure are formulated. Simulation results are presented in Section 3 in order to illustrate the performance of the controller. In Section 4 extensive experimental results are discussed. Section 5 concludes the paper.  

\section{Controller and observer design}

\subsection{Dynamics of a perturbed cascaded system}

Consider a second order fully actuated control system defined
as follows 
\begin{equation}
\left\{ \begin{array}{cl}
\dot{x}_{1} & =x_{2},\\
\dot{x}_{2} & =Bu+h(x_{1},x_{2})+q(x_{1},x_{2},u,t),
\end{array}\right.\label{eq:general:nominal system}
\end{equation}
where $x_{1},\,x_{2}\in\mathbb{R}^{n}$ are state variables, $B\in\mathbb{R}^{n\times n}$
is a non-singular input matrix while $u\in\mathbb{R}^{n}$ stands for
an input. Functions $h:\mathbb{R}^{2n}\rightarrow\mathbb{R}^{n}$ and $q:\mathbb{R}^{2n}\times\mathbb{R}_{\geq 0}\rightarrow\mathbb{R}^{n}$
denote known and unknown components of the dynamics, respectively. Next, it is assumed that input $u$ in \eqref{eq:general:nominal system} is not directly accessible for a control purpose, however, it is governed by the following
first order dynamics 
\begin{equation}
\dot{u}=T^{-1}\left(-u+v\right),\label{eq:general:input dynamics}
\end{equation}
where $v\in\mathbb{R}^{n}$ is regarded as a real input and $T\in\mathbb{R}^{n\times n}$
is a diagonal matrix of positive time constants. In fact, both dynamics constitute a cascaded third order plant, for which the underlying component is represented by \eqref{eq:general:nominal system}, while  \eqref{eq:general:input dynamics} corresponds to  stable input dynamics.

\subsection{Control system design}

The control task investigated in this paper deals with tracking of a reference trajectory specified for an output of system \eqref{eq:general:nominal system}-\eqref{eq:general:input dynamics} which is determined by $y:=x_1$. Simultaneously, it is assumed that variables $x_2$ and $u$ are unavailable for measurement and the only information is provided by the output.  

To be more precise, we define at least $C^3$ continuous reference trajectory $x_{d}(t):\mathbb{R}^{n}\rightarrow\mathbb{R}^{n}$ and consider output tracking error $\tilde{y}:=x_d-x_1$. 
Additionally, to quantify a difference
between $u$ and $v$, we introduce error $\tilde{u}:=v-u$. Since $v$ is viewed as an alternative input of \eqref{eq:general:nominal system}, one can rewrite \eqref{eq:general:nominal system} as
\begin{equation}
\left\{ \begin{array}{cl}
\dot{x}_{1} & =x_{2},\\
\dot{x}_{2} & =Bv-B\tilde{u}+h+q.
\end{array}\right.\label{eq:general:nominal_system_input_v}
\end{equation}
For control design purposes, the tracking error will be considered with respect to the state of system \eqref{eq:general:nominal_system_input_v}. Consequently, one defines
\begin{equation}
e = \begin{bmatrix}e_1\\ e_2\end{bmatrix}:=\begin{bmatrix}\tilde{y}\\ e_2\end{bmatrix}=\begin{bmatrix}x_d-x_1\\ \dot{x}_d-x_2\end{bmatrix}\in\mathbb{R}^{2n}.
\end{equation}
Accordingly, taking time derivative of $e$, one can obtain the following open-loop error dynamics
\begin{equation}
\left\{ \begin{array}{cl}
\dot{e}_{1} & =e_{2},\\
\dot{e}_{2} & =\ddot{x}_{d}-Bv+B\tilde{u}-h-q.
\end{array}\right.\label{eq:general:tracking error dynamics}
\end{equation}
In order to stabilise system \eqref{eq:general:tracking error dynamics} in a vicinity of zero, the following preliminary control law is proposed
\begin{equation}
v:=B^{-1}\left(K_{p}\left(x_{d}-\hat{x}_{1}\right)+K_{d}\left(\dot{x}_{d}-\hat{x}_{2}\right)-h_{u}+\ddot{x}_{d}-w_c\right),\label{eq:general:control law}
\end{equation}
where $K_{p},K_{d}\in\mathbb{R}^{n}$ are diagonal matrices of constant positive gains, $\hat{x}_1\in\mathbb{R}^n$, $\hat{x}_2\in\mathbb{R}^n$ and $w_c\in\mathbb{R}^n$ denote estimates of states and a disturbance, respectively. These estimates are computed by an observer that is not yet defined. Term  $h_{u}:\mathbb{R}^{4n}\rightarrow\mathbb{R}^{n}$
is a compensation function, designed in attempt to attenuate influence of $h$ on the closed system dynamics, and is defined using available signals as
follows
\begin{equation}
h_{u}:=h_{1}(\hat{x}_{1},\hat{x}_{2})+h_{2}(x_{d},\dot{x}_{d}),\label{eq:general:known dynamics compensation}
\end{equation}
while $h_1$ and $h_2$ satisfy 
\begin{equation}
h_{1}(x_{1},x_{2})+h_{2}(x_{1},x_{2}):=h.\label{eq:general:known dynamics}
\end{equation}
Next, in order to simplify design of an observer we rewrite dynamics \eqref{eq:general:nominal system}. Firstly, we consider a new form which does not introduce any change to the system dynamics and is as follows
\begin{equation}
\left\{ \begin{array}{cl}
\dot{x}_{1} & =x_{2},\\
\dot{x}_{2} & =Bu+h_u+h-h_{u}+q.
\end{array}\right.\label{eq:general:nominal system rewritten}
\end{equation}
Secondly, according to active disturbance rejection methodology, it is assumed that
\begin{equation}
z_{3}:=q+h-h_{u}
\end{equation}
describes an augmented state which can be regarded as a total disturbance. Correspondingly, one can introduce extended state $z=\begin{bmatrix}z_{1}^{T} & z_{2}^{T} & z_{3}^{T}\end{bmatrix}^{T}\in\mathbb{R}^{3n}$, where $z_1:=x_1$ and $z_2:=x_2$. As a result, the following extended form of dynamics \eqref{eq:general:nominal system rewritten} can be established
\begin{equation}
\left\{ \begin{array}{cl}
\dot{z}_{1} & =z_{2},\\
\dot{z}_{2} & =Bu+h_{u}+z_{3},\\
\dot{z}_{3} & =\dot{q}+\dot{h}-\dot{h}_{u}.
\end{array}\right.\label{eq:general:extended system-1}
\end{equation}
Now, in order to estimate state $z$ we define the following Luenberger-like observer
\begin{equation}
\left\{ \begin{array}{cl}
\dot{\hat{z}}_{1} & =K_{1}\left(z_{1}-\hat{z}_{1}\right)+\hat{z}_{2},\\
\dot{\hat{z}}_{2} & =K_{2}\left(z_{1}-\hat{z}_{1}\right)+\hat{z}_{3}+h_{u}+Bv,\\
\dot{\hat{z}}_{3} & =K_{3}\left(z_{1}-\hat{z}_{1}\right),
\end{array}\right.\label{eq:general:observer}
\end{equation}
where $\hat{z}=\begin{bmatrix}\hat{z}_{1}^{T} & \hat{z}_{2}^{T} & \hat{z}_{3}^{T}\end{bmatrix}^{T}\in\mathbb{R}^{3n}$ denotes estimate of $z$ and $K_{1},K_{2},K_{3}\in\mathbb{R}^{n\times n}$
are diagonal matrices of positive gains of the observer which are chosen based on linear stability criteria. Since estimates $\hat{z}$ are expected to converge
to real values of $z$, let observation errors be expressed as $\tilde{z}:=z-\hat{z}$. Taking time derivative of $\tilde{z}$, using \eqref{eq:general:observer}, \eqref{eq:general:extended system-1} and recalling \eqref{eq:general:nominal_system_input_v} one obtains the following dynamics 
\begin{equation}
\dot{\tilde{z}}=H_{o}\tilde{z}+C_{0}B\tilde{u}+C_{1}\dot{z}_{3}\label{eq:general:observator error dynamics}
\end{equation}
where
\begin{equation}\label{eq:general:Ho_def}
H_{o}=\begin{bmatrix}-K_{1} & I & 0\\
-K_{2} & 0 & I\\
-K_{3} & 0 & 0
\end{bmatrix}\in\mathbb{R}^{3n\times 3n},
\end{equation}
\begin{equation*}
C_{0}=\begin{bmatrix}0& -I& 0\end{bmatrix}^T,\ C_{1}=\begin{bmatrix}0& 0& I
\end{bmatrix}^T\in\mathbb{R}^{3n},
\end{equation*}
while $I$ stands for the identity matrix of size $n\times n$. Here, it is required that $H_{o}$ is Hurwitz, what can be guaranteed by a proper choice of observer gains. 
Next, we recall tracking dynamics \eqref{eq:general:tracking error dynamics} and feedback \eqref{eq:general:control law}. It is proposed that compensating term in \eqref{eq:general:control law}, which partially rejects unknown disturbances, is defined by an estimate provided by observer \eqref{eq:general:observer}, namely  $w_c:=\hat{z}_3$. Consequently, by substituting
\eqref{eq:general:control law} into \eqref{eq:general:tracking error dynamics}
the following is obtained
\begin{equation}
\dot{e}=H_{c}e+W_{1}\tilde{z}+C_{2}B\tilde{u},\label{eq:general:regulator error dynamics}
\end{equation}
where 
\begin{equation}\label{eq:general:Hc_def}
H_{c}=\begin{bmatrix}0 & I\\
-K_{p} & -K_{d}
\end{bmatrix},W_{1}=\begin{bmatrix}0 & 0 & 0\\
-K_{p} & -K_{d} & -I
\end{bmatrix},\ C_{2}=\begin{bmatrix}0\\
I
\end{bmatrix}\in\mathbb{R}^{2n\times n}
\end{equation}
and $H_{c}$ is Hurwitz for $K_{p}\succ 0$ and $K_{d}\succ 0$.

Further, in order to facilitate the design and analysis of the closed-loop system, we take advantage of a scaling operator defined by
\begin{equation}
\Delta_m\left(\alpha\right):=\mathrm{diag}\left\{\alpha^{m-1}I,\, \alpha^{m-2}I,\, \ldots,\, I\right\}\in\mathbb{R}^{mn\times mn},
\end{equation}
where $\alpha>0$ is a positive scalar. Then we define the following scaled tracking and observation errors
\begin{align}
\bar{e}:=&\left(\kappa\omega\right)^{-1}\Delta_2\left(\kappa\omega\right)e,\label{eq:general:regulator auxiliary errors}\\
\bar{z}:=&\omega^{-2}\Delta_3\left(\omega\right)\tilde{z},\label{eq:general:observer auxiliary errors} 
\end{align}
where $\omega\in\mathbb{R}_{+}$ is scaling parameter which modifies the bandwidth of the the observer, while $\kappa\in\mathbb{R}_{+}$ denotes a relative bandwidth of the feedback determined with respect to $\omega$. Embracing this notation one can introduce the following scaled gains 
\begin{equation}\label{eq:design:scaled_gains}
\bar{K}_c:=\left(\kappa\omega\right)^{-1}K_c \Delta_2^{-1}\left(\kappa\omega\right),\, \bar{K}_o:=\omega^{-3}\Delta_3\left(\omega\right) \left[K_1^T\ K_2^T\ K_3^T\right]^T,
\end{equation}
while $K_c:=\left[K_p\ K_d\right]\in\mathbb{R}^{n\times 2n}$. Additionally, exploring relationships \eqref{eq:app:scalled_terms} outlined in the Appendix, one can rewrite dynamics \eqref{eq:general:regulator error dynamics} and \eqref{eq:general:observator error dynamics} as follows
\begin{align}
\dot{\bar{e}}=&\kappa\omega\bar{H}_{c}\bar{e}+\kappa^{-1}\omega\bar{W}_{1}\Delta_3\left(\kappa\right)\bar{z}+\left(\kappa\omega\right)^{-1}C_{2}B\tilde{u},\label{eq:general:regulator auxiliary error dynamics}\\
\dot{\bar{z}}=&\omega\bar{H}_{o}\bar{z}+\omega^{-1} C_{0}B\tilde{u}+\omega^{-2} C_{1}\dot{z}_{3},\label{eq:general:observator auxiliary error dynamics}
\end{align}
with $\bar{H}_c$ and $\bar{H}_o$ being Hurwitz matrices of forms \eqref{eq:general:Hc_def}, \eqref{eq:general:Ho_def} defined in terms of scaled gains $\bar{K}_c$ and $\bar{K}_o$, respectively. Similarly, $\bar{W}_1$ corresponds to $W_1$ parameterised by new gains. 
Since $\bar{H}_c$ and $\bar{H}_o$ are Hurwitz, one can state that the following Lyapunov equations are satisfied
\begin{equation}
\bar{P}_c\bar{H}_c^{T}+\bar{H}_c\bar{P}_c=-\bar{Q}_c,\ \bar{P}_o\bar{H}_o^{T}+\bar{H}_o\bar{P}_o=-\bar{Q}_o \label{eq:general:Lyapunov equation}
\end{equation}
for some symmetric, positive defined matrices $\bar{Q}_c,\, \bar{P}_c\in\mathbb{R}^{2n\times 2n}$ and $\bar{Q}_o,\, \bar{P}_o\in\mathbb{R}^{3n\times 3n}$.

\subsection{Stability analysis of the closed-loop cascaded control system}

Lyapunov stability of the closed-loop is
to be considered now. For this purpose, a state which consists of tracking, observation and input errors is defined as 
\begin{equation}
\bar\zeta=\begin{bmatrix}\bar{e}^T&\bar{z}^T&\tilde{u}^T\end{bmatrix}^T\in\mathbb{R}^{6n}.\label{eq:general:stability:errors}
\end{equation}
A positive definite function is proposed as follows 
\begin{equation}
V(\bar{\zeta})=\frac{1}{2}\bar{e}^{T}\bar{P}_c\bar{e}+\frac{1}{2}\bar{z}^{T}\bar{P}_o\bar{z}+\frac{1}{2}\tilde{u}^{T}\tilde{u}.\label{eq:general:stability:lyapunov proposition}
\end{equation}
Its derivative takes form of 
\begin{equation}
\begin{aligned}
\dot{V}(\bar{\zeta})=&-\frac{1}{2}\kappa\omega\bar{e}^{T}\bar{Q}_c\bar{e}-\frac{1}{2}\omega\bar{z}^{T}\bar{Q}_o\bar{z}+\kappa^{-1}\omega \bar{e}^T \bar{P}_c\bar{W}_1\Delta_3\left(\kappa\right)\bar{z}+\left(\kappa\omega\right)^{-1}\bar{e}^{T}\bar{P}_c C_2B\tilde{u}+\omega^{-1}\bar{z}^T\bar{P}_o C_o B\tilde{u}\\&+\omega^{-2}\bar{z}^{T}\bar{P}_o C_1\dot{z}_{3}-\tilde{u}^{T}T^{-1}\tilde{u}+\tilde{u}^{T}\dot{v}.\label{eq:general:stability:lyapunov derivative}
\end{aligned}
\end{equation}
Derivative of control law $v$ defined by \eqref{eq:general:control law} can be expressed in terms of $\bar{\zeta}$
as (the details are outlined in the Appendix)
\begin{equation}
\dot{v}=B^{-1}\left(\omega^{3}\left(\kappa^3\bar{K}_c\bar{H}_c\bar{e}+\left(\kappa\bar{K}_c\bar{W}_1 \Delta_3\left(\kappa\right)+\bar{W}_2\Delta_3\left(\kappa\right)\bar{H}_o\right)\bar{z}\right)-\dot{h}_u+\dddot{x}_{d}\right),\label{eq:general:stability:control law derivative}
\end{equation}
where $\bar{K}_c:=\left[\bar{K}_p\ \bar{K}_d\right]\in\mathbb{R}^{n\times 2n}$ and $\bar{W}_2:=\left[\bar{K}_c\ I \right]\in\mathbb{R}^{n\times{3n}}$. Substituting (\ref{eq:general:stability:control law derivative})
and $\dot{z}_{3}$ into (\ref{eq:general:stability:lyapunov derivative})
leads to
\begin{equation}
\begin{aligned}
\dot{V}(\bar{\zeta})=&-\frac{1}{2}\kappa\omega\bar{e}^{T}\bar{Q}_c\bar{e}-\frac{1}{2}\omega\bar{z}^{T}\bar{Q}_o\bar{z}+\kappa^{-1}\omega\bar{e}^T\bar{P}_c\bar{W}_{1}\Delta_3\left(\kappa\right)\bar{z}+\left(\kappa\omega\right)^{-1}\bar{e}^{T}\bar{P}_c C_2B\tilde{u}+\omega^{-1}\bar{z}^T\bar{P}_o C_o B\tilde{u}\\&+\left(\kappa\omega\right)^3 \tilde{u}^{T}B^{-1}\bar{K}_c\bar{H}_c\bar{e}+\omega^3\tilde{u}^{T}B^{-1}\left(\kappa\bar{K}_c\bar{W}_1 \Delta_3\left(\kappa\right)+\bar{W}_2\Delta_3\left(\kappa\right)\bar{H}_o\right)\bar{z}-\tilde{u}^{T}T^{-1}\tilde{u}\\ &+\tilde{u}^{T}B^{-1}\dddot{x}_d+\tilde{u}^{T}B^{-1}\dot{h}_u+\omega^{-2}\bar{z}^{T}\bar{P}_o C_1\left(\dot h-\dot{h}_u\right)+\omega^{-2}\bar{z}^{T}\bar{P}_o C_1\dot{q}(z_{1},z_{2},u,t).\label{eq:general:stability:lyapunov derivative split}
\end{aligned}
\end{equation}
In order to simplify the stability analysis, derivative $\dot{V}$ will be decomposed into four terms defined as follows
\begin{equation}
\begin{aligned}
Y_1:=&-\frac{1}{2}\kappa\omega\bar{e}^{T}\bar{Q}_c\bar{e}-\frac{1}{2}\omega\bar{z}^{T}\bar{Q}_o\bar{z}+\kappa^{-1}\omega\bar{e}^T\bar{P}_c\bar{W}_{1}\Delta_3\left(\kappa\right)\bar{z}+\left(\kappa\omega\right)^{-1}\bar{e}^{T}\bar{P}_c C_2B\tilde{u}+\omega^{-1}\bar{z}^T\bar{P}_o C_o B\tilde{u}\\&+\left(\kappa\omega\right)^3 \tilde{u}^{T}B^{-1}\bar{K}_c\bar{H}_c\bar{e}+\omega^3\tilde{u}^{T}B^{-1}\left(\kappa\bar{K}_c\bar{W}_1 \Delta_3\left(\kappa\right)+\bar{W}_2\Delta_3\left(\kappa\right)\bar{H}_o\right)\bar{z}-\tilde{u}^{T}T^{-1}\tilde{u},\\
Y_2:=& \tilde{u}^{T}B^{-1}\dddot{x}_d,\,Y_3:=\tilde{u}^{T}B^{-1}\dot{h}_u+\omega^{-2}\bar{z}^{T}\bar{P}_o C_1\left(\dot h-\dot{h}_u\right),\,Y_4:= \omega^{-2}\bar{z}^{T}\bar{P}_o C_1\dot{q}(z_{1},z_{2},u,t).
\end{aligned}
\end{equation}
Each term of $\dot{V}$ will be now considered separately. Firstly,
$Y_{1}$ which represents mainly influence of input dynamics on the nominal
system will be looked upon. Negative definiteness of this term will
be a starting point for further analysis of the closed-loop stability.
Let it be rewritten using the matrix notation as
\begin{equation}
Y_{1} =-\frac{1}{2}\omega\bar{\zeta}^{T}Q_{Y1}\bar{\zeta},\label{eq:general:stability:Y1}
\end{equation}
where 
\[
\begin{split}
Q_{Y1}=\left[\begin{matrix}\kappa\bar{Q}_c &-\kappa^{-1}\bar{P}_c\bar{W}_1\Delta_3\left(\kappa\right)&Q_{Y1_{13}}\\-\kappa^{-1}\left(\bar{P}_c\bar{W}_1\Delta_3\left(\kappa\right)\right)^T&\bar{Q}_o&Q_{Y1_{23}}\\
Q_{Y1_{13}}^T&Q_{Y1_{23}}^T&2\omega^{-1} T^{-1}
\end{matrix}\right]\in\mathbb{R}^{6n\times 6n}
\end{split}
\]
while
\begin{equation}
\begin{aligned}
Q_{Y1_{13}} =& -\kappa^{-1}\omega^{-2}\bar{P}_c C_2B-\kappa^3\omega^2\left(B^{-1}\bar{K}_c\bar{H}_c\right)^T,\\
Q_{Y1_{23}}=&-\omega^{-2}\bar{P}_o C_o B-\omega^2\left(B^{-1}\left(\kappa\bar{K}_c\bar{W}_1 \Delta_3\left(\kappa\right)+\bar{W}_2\Delta_3\left(\kappa\right)\bar{H}_o\right)\right)^T.
\end{aligned}
\end{equation}
It can be showed, that there may exist sets $\Omega_v, \mathrm{K}_v \subset \mathbb{R}_{+}$, such, that for every $\omega\in\Omega_v$ and $\kappa\in\mathrm{K}_v$ matrix $Q_{Y1}$ remains positive definite. Domains of both $\Omega_v$ and $\mathrm{K}_v$ strongly depend on inertia matrix $T$ and input matrix $B$ of nominal system. In the absence of other disturbances system would remain asymptotically stable for such a choice of both $\omega$ and $\kappa$ parameters. Influence of other elements of $\dot{V}(\bar{\zeta})$
will be considered in terms of upper bounds which can be imposed on
them.
\begin{assumption}
\label{assu:desired trajectory}Let desired trajectory $x_{d}$ be
chosen such, that norms of $x_{d},\dot{x}_{d},\ddot{x}_{d},\dddot{x}_{d}$
are bounded by, respectively, constant positive scalar values $x_{b0},x_{b1},x_{b2},x_{b3}\in\mathbb{R}_{+}$.
\end{assumption}
Establishing upper bound for norm of $Y_{2}$ is straightforward by
using Cauchy-Schwartz inequality. 
\begin{align}
Y_{2} & =-\tilde{u}^{T}B^{-1}\dddot{x}_{d},\nonumber \\
\left\Vert Y_{2}\right\Vert  & \leq\left\Vert \tilde{u}\right\Vert \cdot\left\Vert B^{-1}\dddot{x}_{d}\right\Vert \nonumber \\
 & \leq\left\Vert \bar{\zeta}\right\Vert \left\Vert B^{-1}\right\Vert x_{b3}.\label{eq:general:stability:Y2}
\end{align}
Now, $Y_{3}$ is to be considered. This term comes from imperfect
compensation of known dynamics in nominal system and it can be further split into the following
\begin{equation}
Y_{31}:=\omega^{-2}\bar{z}^{T}P_{o}C_{1}\left(\dot{h}-\dot{h}_{u}\right), Y_{32}:=\tilde{u}^{T}B^{-1}\dot{h}_{u}.\label{eq:general:stability:Y3}
\end{equation}

\begin{assumption}
\label{assu:bounded dynamics}Let functions $h_{1}(a,b)$ and $h_{2}(a,b)$
be defined such, that norms of partial derivatives\\
 $\frac{\partial}{\partial a}h_{1}(a,b)$,$\frac{\partial}{\partial b}h_{1}(a,b)$,$\frac{\partial}{\partial a}h_{2}(a,b)$,$\frac{\partial}{\partial b}h_{2}(a,b)$
are bounded for every $a,b\in\mathbb{R}^{n}$ by $h_{1a},h_{1b,}h_{2a,}h_{2b}\in\mathbb{R}_{+}$
respectively.
\end{assumption}
By applying chain rule to calculate derivatives of each function and
substituting difference of error and desired trajectory for state
variables, term $Y_{31}$ can be expressed as 
\begin{equation}
Y_{31}=\omega^{-2}\bar{z}^{T}P_{o}C_{1}\left(W_{h1}\begin{bmatrix}\dot{x}_{d}\\\ddot{x}_{d}\end{bmatrix}
- W_{h2}\left(\kappa\omega\bar{H}_{c}\bar{e}+\kappa^{-1}\omega\bar{W}_{1}\Delta_3(\kappa)\bar{z}+\left(\kappa\omega\right)^{-1}C_{2}B\tilde{u}\right)+W_{h3}\left(\omega\bar{H}_{o}\bar{z}+\omega^{-1}C_{0}B\tilde{u}\right)\right),\label{eq:general:stability:Y31 equation}
\end{equation}
where 
\begin{align}
W_{h1} & =\begin{bmatrix}\left(\frac{\partial h_{1}}{\partial z_{1}}+\frac{\partial h_{2}}{\partial z_{1}}-\frac{\partial h_{2}}{\partial x_{d}}-\frac{\partial h_{1}}{\partial\hat{z}_{1}}\right) & \left(\frac{\partial h_{1}}{\partial z_{2}}+\frac{\partial h_{2}}{\partial z_{2}}-\frac{\partial h_{2}}{\partial\dot{x}_{d}}-\frac{\partial h_{1}}{\partial\hat{z}_{2}}\right)\end{bmatrix}, \nonumber \\
W_{h2} & =\begin{bmatrix}\left(\frac{\partial h_{1}}{\partial z_{1}}+\frac{\partial h_{2}}{\partial z_{1}}-\frac{\partial h_{1}}{\partial\hat{z}_{1}}\right) & \kappa\omega\left(\frac{\partial h_{1}}{\partial z_{2}}+\frac{\partial h_{2}}{\partial z_{2}}-\frac{\partial h_{1}}{\partial\hat{z}_{2}}\right)\end{bmatrix}, \nonumber \\
W_{h3} & =\begin{bmatrix}\frac{\partial h_{1}}{\partial\hat{z}_{1}} & \omega\frac{\partial h_{1}}{\partial\hat{z}_{2}} & 0\end{bmatrix}. \nonumber
\end{align}
This term can be said to be bounded by 
\begin{align}
\left\Vert Y_{31}\right\Vert \leq & \omega^{-2}\left\Vert \bar{\zeta}\right\Vert \left\Vert P_{o}C_{1}\right\Vert \left(\left(2h_{1a}+2h_{2a}\right)x_{b1}+\left(2h_{1b}+2h_{2b}\right)x_{b2}\right) \nonumber \\
& +\left\Vert \bar{\zeta}\right\Vert ^{2}\left\Vert P_{o}C_{1}\right\Vert \left(\omega^{-1}\kappa\left\Vert W_{h2b}\right\Vert \left(\left\Vert \bar{H}_{c}\right\Vert +\left\Vert \bar{W}_{1}\right\Vert \right)+\omega^{-3}\kappa^{-1}\left\Vert W_{h2b}\right\Vert \left\Vert C_{2}B\right\Vert\right) \nonumber \\
& +\left\Vert \bar{\zeta}\right\Vert ^{2}\left\Vert P_{o}C_{1}\right\Vert \left(\omega^{-1}\left\Vert W_{h3b}\right\Vert \left\Vert \bar{H}_{o}\right\Vert +\omega^{-2}\left\Vert B\right\Vert h_{1b} \right). \label{eq:general:stability:Y31 bound}
\end{align}
where $W_{h2b}=\begin{bmatrix}2h_{1a} + h_{2a} & \kappa\omega\left(2h_{1b} + h_{2b}\right)\end{bmatrix}$ and $W_{h3b} = \begin{bmatrix}h_{1a} & \omega h_{1b} & 0 \end{bmatrix}$.
Having established upper bound of $Y_{31}$, we can perform similar analysis with respect to $Y_{32}$. Let $Y_{32}$ be rewritten as
\begin{equation}
Y_{32} = \tilde{u}^T B^{-1} \left(W_{h4}\begin{bmatrix}\dot{x}_{d}\\\ddot{x}_{d}\end{bmatrix}
- W_{h5}\left(\kappa\omega\bar{H}_{c}\bar{e}+\kappa^{-1}\omega\bar{W}_{1}\Delta_3(\kappa)\bar{z}+\left(\kappa\omega\right)^{-1}C_{2}B\tilde{u}\right)-W_{h6}\left(\omega\bar{H}_{o}\bar{z}+\omega^{-1}C_{0}B\tilde{u}\right)\right), \label{eq:general:stability:Y32 equation}
\end{equation}
where
\begin{align}
W_{h4} & =\begin{bmatrix}\left(\frac{\partial h_{2}}{\partial x_{d}}+\frac{\partial h_{1}}{\partial \hat{z}_{1}}\right) & \left(\frac{\partial h_{2}}{\partial \dot{x}_{d}}+\frac{\partial h_{1}}{\partial \hat{z}_{2}}\right)\end{bmatrix}, \nonumber \\
W_{h5} & =\begin{bmatrix}\frac{\partial h_{1}}{\partial \hat{z}_{1}} & \kappa\omega\frac{\partial h_{1}}{\partial \hat{z}_{2}}\end{bmatrix}, \nonumber \\
W_{h6} & = W_{h3}. \nonumber    
\end{align}
An upper bound of norm of $Y_{32}$ can be expressed by the following inequality
\begin{align}
\left\Vert Y_{32} \right\Vert \leq & \omega^{-2}\left\Vert \bar{\zeta}\right\Vert \left\Vert \bar{P}_{o}C_{1}\right\Vert \left(q_{z1}x_{b1}+q_{z2}x_{b2}+\left\Vert B\right\Vert q_{z2}+\left\Vert T^{-1}\right\Vert q_{u}+\left\Vert \bar{P}_{o}C_{1}\right\Vert q_{t}\right) \nonumber \\
& +\kappa\omega^{-1}\left\Vert \bar{\zeta}\right\Vert ^{2}\left\Vert \bar{P}_{o}C_{1}\right\Vert \left\Vert W_{q2}\right\Vert \left(\left\Vert \bar{H}_{c}\right\Vert +\left\Vert \bar{W}_{1}\right\Vert \right)\label{eq:general:stability:Y_32 bound}
\end{align}
where $W_{h5b} = \begin{bmatrix}h_{1a} & \kappa\omega h_{1b}\end{bmatrix}$ and naturally $W_{h6b}=W_{h3b}$. A remark can be made now about the structure of $W_{h2}$, $W_{h3}$, $W_{h4}$ and $W_{h5}$. It may be recognized, that elements of these matrices can be divided into group of derivatives calculated with respect to the first and the second argument. Former of these are not scaled by either observer or regulator bandwidth, while the latter is scaled by either $\kappa\omega$ or $\omega$ factor. As will be showed later in the analysis, this difference will have significant influence on the system stability and ability of the controller to reduce tracking errors. 

Lastly, some upper bound need to be defined for $Y_{4}$ to complete the
stability analysis. This final term comes from nominal disturbance
$q(z_1, z_2, u, y)$ alone. By chain rule it can be shown that 
\begin{equation}
Y_{4} = \omega^{-2}\bar{z}^{T}\bar{P}_{o}C_{1}\left(W_{q1}\begin{bmatrix}\dot{x}_{d}\\\ddot{x}_{d}\end{bmatrix}+\kappa\omega W_{q2}\bar{H}_{c}\bar{e}+\kappa^{-1}\omega W_{q2}\bar{W}_{1}\Delta_3\left(\kappa\right)\bar{z}+\left(\kappa\omega\right)^{-1}W_{q2}C_{2}B\tilde{u}-\frac{\partial q}{\partial u}T^{-1}\tilde{u}+\frac{\partial q}{\partial t}\right),\label{eq:general:stability:Y4}
\end{equation}
where $W_{q1}=\begin{bmatrix}\frac{\partial q}{\partial z_1} & \frac{\partial q}{\partial z_2}\end{bmatrix}$ and $W_{q2} = \begin{bmatrix}\frac{\partial q}{\partial z_1} & \kappa\omega\frac{\partial q}{\partial z_2}\end{bmatrix}$.
\begin{assumption}
\label{assu:disturbance derivatives}Let partial derivatives $\frac{\partial}{\partial z_{1}}q(z_{1},z_{2},u,t),\frac{\partial}{\partial z_{2}}q(z_{1},z_{2},u,t),\frac{\partial}{\partial u}q(z_{1},z_{2},u,t),\frac{\partial}{\partial t}q(z_{1},z_{2},u,t)$
be defined in the whole domain and let their norms be bounded by constants
$q_{z1},q_{z2},q_{u}$ and $q_{t}\in\mathbb{R}_{+}$, respectively.
\end{assumption}
Under Assumption \ref{assu:disturbance derivatives} the norm of
$Y_{4}$ is bounded by 
\begin{align}
\left\Vert Y_{4}\right\Vert  & \leq\omega^{-2}\left\Vert \bar{\zeta}\right\Vert \left\Vert \bar{P}C\right\Vert \left(q_{z1}x_{b1}+q_{z2}x_{b2}\right)+\omega^{-1}\left\Vert \bar{\zeta}\right\Vert ^{2}\left\Vert \bar{P}CW_{5b}\right\Vert \left(\left\Vert \bar{H}\right\Vert +\omega^{-2}\left\Vert \bar{C}B\right\Vert \right)\label{eq:general:stability:Y4 bound}\\
 & +\omega^{-2}\left\Vert \bar{\zeta}\right\Vert ^{2}\left\Vert \bar{P}C\right\Vert \left(\left\Vert T^{-1}\right\Vert q_{u}+q_{t}\right).\nonumber 
\end{align}
With some general bounds for each of $\dot{V}(\bar{\zeta})$
terms established, conclusions concerning system stability can be finally drawn. For the sake of convenience, let some auxiliary measure of Lyapunov function derivative negative definiteness $\Lambda_V$ and Lyapunov function derivative perturbation $\Gamma_V$ be defined as
\begin{align}
\Lambda_V := & \frac{1}{2}\omega\lambda_{\min}(Q_{Y1}) -\kappa\omega\left\Vert B^{-1}\right\Vert \left\Vert W_{h5b}\right\Vert \left(\left\Vert \bar{H}_{c}\right\Vert +\left\Vert \bar{W}_{1}\right\Vert \right)-\omega\left\Vert B^{-1}\right\Vert \left\Vert W_{h6b}\right\Vert \left\Vert \bar{H}_{o}\right\Vert \nonumber \\
& -2h_{1b}-\omega^{-1}\left\Vert W_{h3b}\right\Vert \left\Vert \bar{H}_{o}\right\Vert -\kappa\omega^{-1}\left\Vert P_{o}C_{1}\right\Vert \left(\left\Vert W_{h2b}\right\Vert +\left\Vert W_{q2}\right\Vert \right)\left(\left\Vert \bar{H}_{c}\right\Vert +\left\Vert \bar{W}_{1}\right\Vert \right) \nonumber \\
& -\omega^{-2}\left\Vert B\right\Vert h_{1b}-\omega^{-3}\kappa^{-1}\left\Vert W_{h2b}\right\Vert \left\Vert C_{2}B\right\Vert, \label{eq:general:stability:Lyapunov negative definiteness} \\
\Gamma_V := & \left\Vert B^{-1}\right\Vert \left\Vert \left(h_{2a}+h_{1a}\right)x_{b1}+\left(h_{2b}+h_{1b}\right)x_{b2}\right\Vert \nonumber \\
& \omega^{-2}\left\Vert \bar{P}_{o}C_{1}\right\Vert \left(q_{z1}x_{b1}+q_{z2}x_{b2}+\left\Vert B\right\Vert q_{z2}+\left\Vert T^{-1}\right\Vert q_{u}+\left\Vert \bar{P}_{o}C_{1}\right\Vert q_{t}\right), \label{eq:general:stability:Lyapunov perturbation}
\end{align}
where $\lambda_{\min}(Q)$ stands for the smallest eigenvalue of matrix $Q$, then upper bound of $\dot{V}_{\bar\zeta}(\bar\zeta)$ can be expressed as 
\begin{equation}
    \dot{V}_{\bar\zeta} \leq -\Lambda_V\left\Vert \bar\zeta \right\Vert^2 + \Gamma_V\left\Vert \bar\zeta \right\Vert. \label{eq:general:stability:Lyapunov derivative bound}
\end{equation}
Now, following conditions can be declared 
\begin{enumerate}[label=\textbf{C\arabic*}]
\item \label{enum:general:stability:condition 1} $\omega\in\Omega_v, \kappa\in\mathrm{K}_v$, 
\item \label{enum:general:stability:condition 2} $\Gamma_V \geq 0$,
\end{enumerate}
and succeeding theorem concludes presented analysis.
\begin{proposition}
Perturbed cascade system (\ref{eq:general:nominal system})-(\ref{eq:general:input dynamics}) satisfying Assumptions \ref{assu:desired trajectory}-\ref{assu:disturbance derivatives}, controlled by feedback (\ref{eq:general:control law}) which is supported by extended state observer (\ref{eq:general:observer}), remains practically stable if there exist symmetric, positive defined matrices $Q_o$ and $Q_c$ such, that conditions \ref{enum:general:stability:condition 1} and \ref{enum:general:stability:condition 2}  can be simultaneously satisfied. Scaled tracking errors $\bar\zeta$ are then bounded as follows
\begin{equation}\label{eq:control:conclusion}
\lim_{t\rightarrow\infty}\left\Vert \bar{\zeta}(t)\right\Vert\leq  \frac{\Gamma_{V}}{\Lambda_{V}}.
\end{equation}
\end{proposition}

\begin{remark}
Foregoing proposition remains valid only if Assumptions \ref{assu:desired trajectory}-\ref{assu:disturbance derivatives} are satisfied. While Assumption \ref{assu:desired trajectory} considers desired trajectory only and can be easily fulfilled for any system with state $x_1$ defined on $\mathbb{R}^n$, a closer look at the remaining assumptions ought to be taken now. Similar in their nature, both concern imperfectly known parts of the system dynamics, with the difference being whether an attempt to implicitly compensate these dynamics is taken or not. As a known dynamic term satisfying Assumption \ref{assu:bounded dynamics} can also be treated as an unknown disturbance, without a loss of generality, only Assumption \ref{assu:disturbance derivatives} has to be commented here.  It can be noted, that for many commonly considered systems this assumption cannot be satisfied. A mechanical system equipped with revolute kinematic pairs can be an example of such system, which dynamics, due to Coriolis and centrifugal forces, have neither bounded time derivative nor bounded partial derivative calculated with respect to second state variable. Engineering practice shows nonetheless that for systems, in which cross-coupling is insignificant enough due to a proper mass distribution, this assumption can be approximately satisfied, at least in a bounded set of the state-space, and the stability analysis holds. The requirement that partial derivatives of any disturbance in the system should be bounded is restrictive one, yet less conservative than commonly used in the ADRC analysis expectation of time derivative boundedness. In this sense, the presented analysis is more liberal than ones considered in the literature and it can be expected that enforced assumptions can be better justified.
\end{remark}

\section{Numerical simulations}

In attempt to further research behaviour of the system in the presence
of unmodelled dynamics governing the input signal numerical simulations
have been conducted. Model of the system has been implemented in
Matlab-Simulink environment. The second order, single degree
of freedom system and the first order dynamics of the input have been modelled according
to the following equations
\begin{equation}
\left\{ \begin{array}{cl}
\dot{x}_{1} & =x_{2},\\
\dot{x}_{2} & =u,
\end{array}\right.\label{eq:simulation:system}
\end{equation}
where
\begin{equation}
\dot{u}=\frac{1}{T}\left(-u+v\right)\label{eq:simulation:input}
\end{equation}
and $v$ is a controllable input of the system. Parameters $T$ and
$\omega$ of the controller were modified in simulations to
investigate how they affect the closed-loop system stability and the tracking accuracy. Chosen parameters of the system are presented in
the table \ref{tab:simulation:gains}. Desired trajectory $x_d$ was selected
as a sine wave with unitary amplitude and frequency of $\unit[\frac{10}{2\pi}]{Hz}$.

\begin{table}[h]
\begin{centering}
\begin{tabular}{|c|c|c|c|c|c|}
\hline
$\bar{K}_{1}$ & $\bar{K}_{2}$ & $\bar{K}_{3}$ & $\bar{K}_{p}$ & $\bar{K}_{d}$& $\kappa$\tabularnewline
\hline 
$3$ & $3$ & $1$ & $1$ & $2$ & $0.01$\tabularnewline
\hline 
\end{tabular}
\par\end{centering}
\caption{Auxiliary gains of the observer and controller\label{tab:simulation:gains}}
\end{table}

\begin{figure}[h]
\centering
\begin{minipage}[t]{0.45\columnwidth}%
\subfloat[$\omega=0.1$]{\includegraphics[width=1\columnwidth]{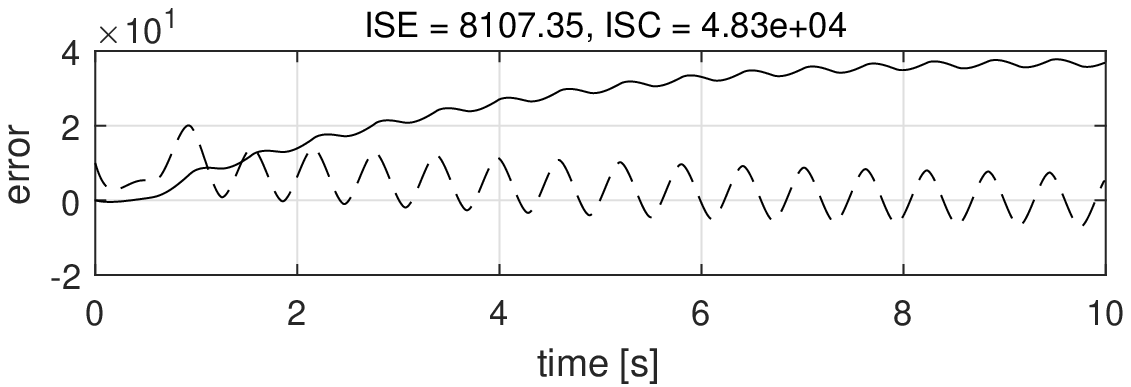}}\par
\subfloat[$\omega=1$]{\includegraphics[width=1\columnwidth]{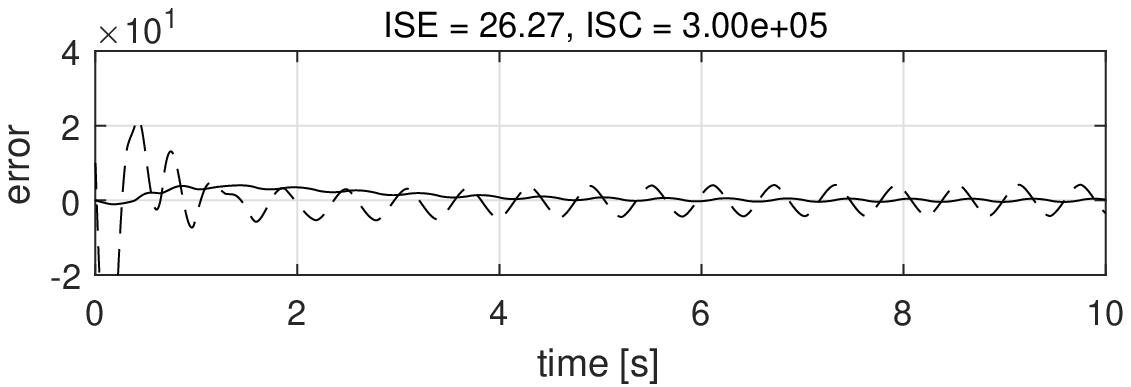}}\par
\subfloat[$\omega=4$]{\includegraphics[width=1\columnwidth]{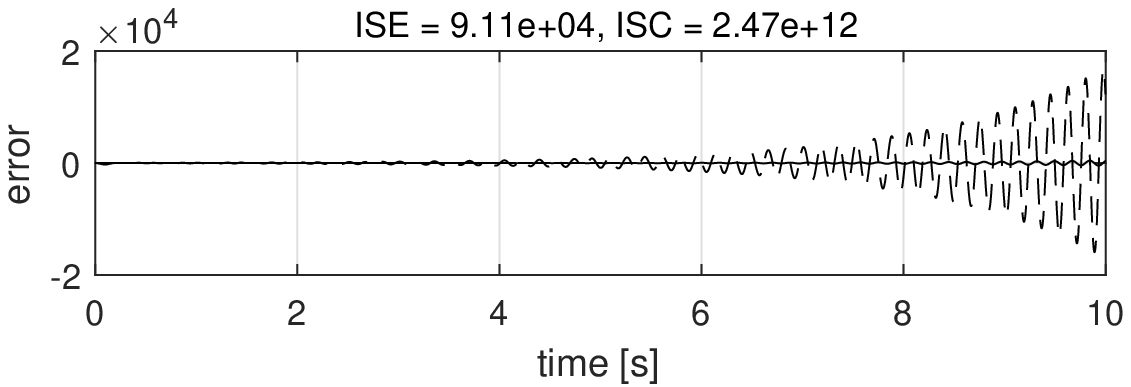}}\par
\caption{$T=\unit[0.1]{s}$, $\hat{z}_{3}$ enabled\label{fig:simulation:T01 adrc}}
\end{minipage}%
\begin{minipage}[t]{0.45\columnwidth}%
\subfloat[$\omega=0.1$]{\includegraphics[width=1\columnwidth]{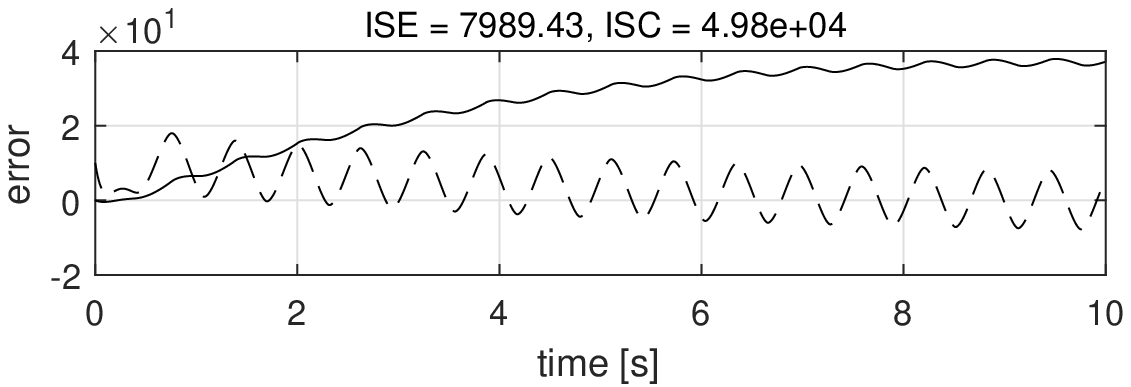}}\par
\subfloat[$\omega=1$]{\includegraphics[width=1\columnwidth]{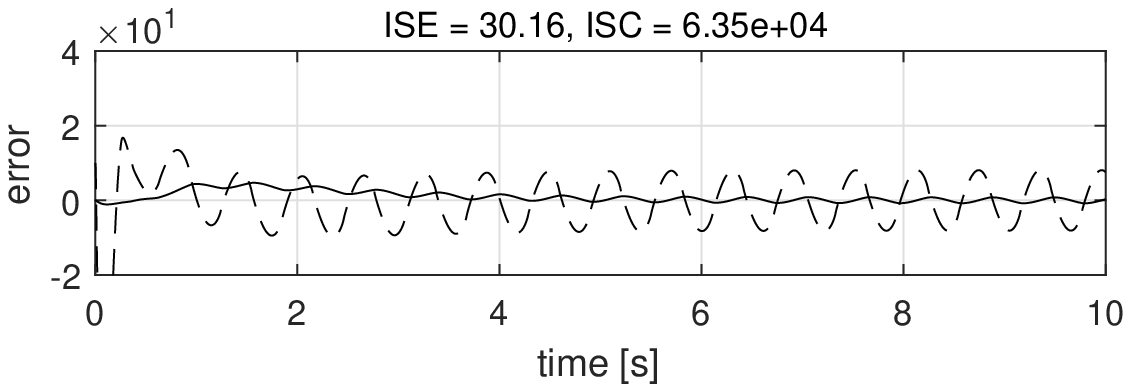}}\par
\subfloat[$\omega=4$]{\includegraphics[width=1\columnwidth]{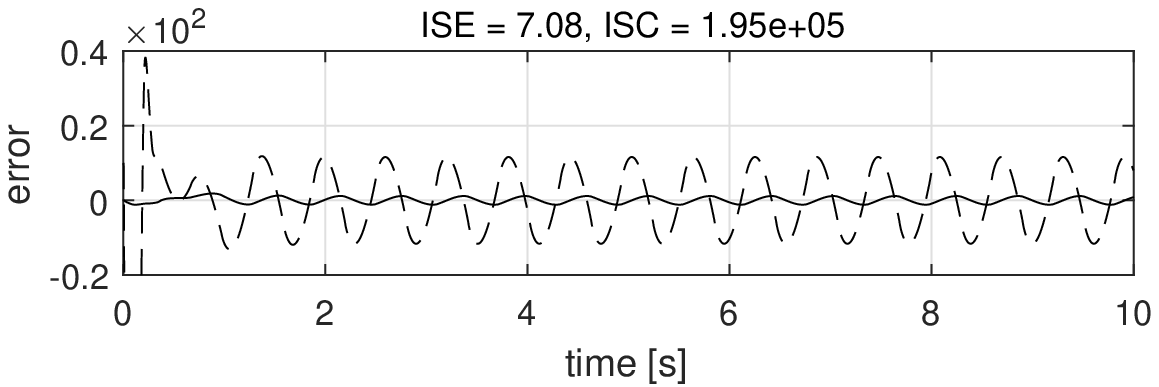}}\par
\caption{$T=\unit[0.1]{s}$, $\hat{z}_{3}$ disabled\label{fig:simulation:T01 pd}}
\end{minipage}
\end{figure}

\begin{figure}[h]
\centering
\begin{minipage}[t]{0.45\columnwidth}%
\subfloat[$\omega=0.1$]{\includegraphics[width=1\columnwidth]{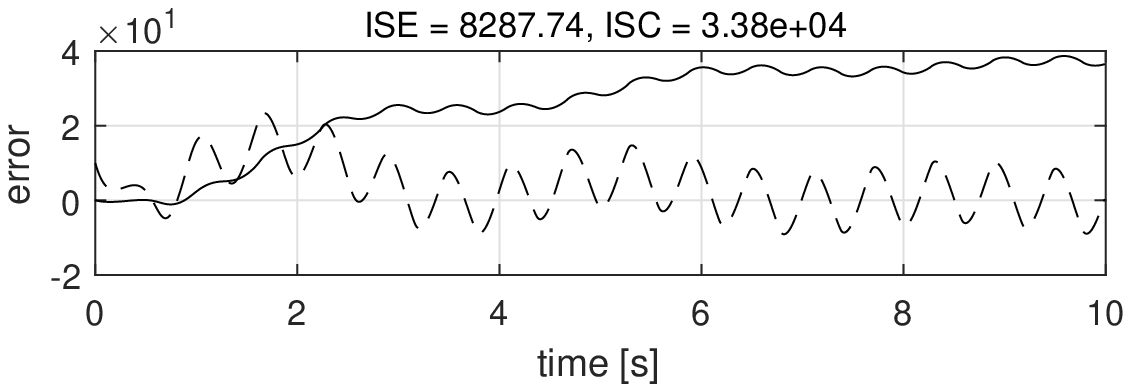}}\par
\subfloat[$\omega=1$]{\includegraphics[width=1\columnwidth]{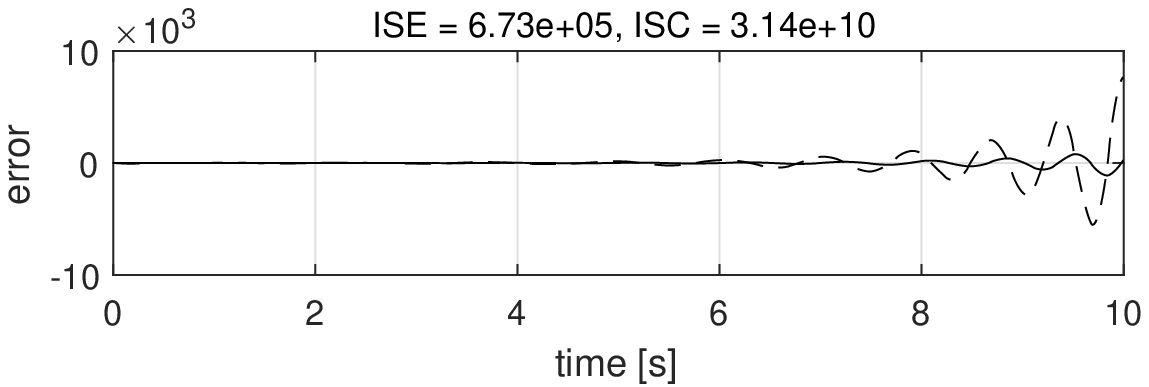}}\par
\subfloat[$\omega=4$]{\includegraphics[width=1\columnwidth]{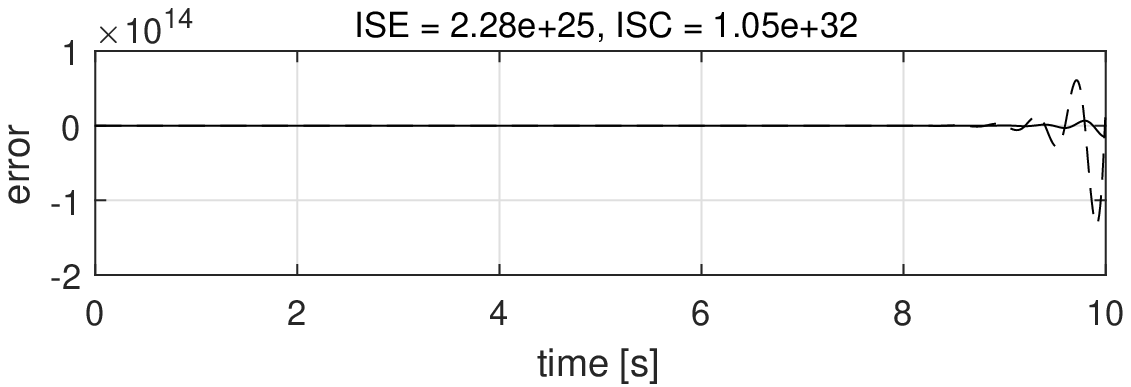}}\par
\caption{$T=\unit[1]{s}$, $\hat{z}_{3}$ enabled\label{fig:simulation:T1 adrc}}
\end{minipage}%
\begin{minipage}[t]{0.45\columnwidth}%
\subfloat[$\omega=0.1$]{\includegraphics[width=1\columnwidth]{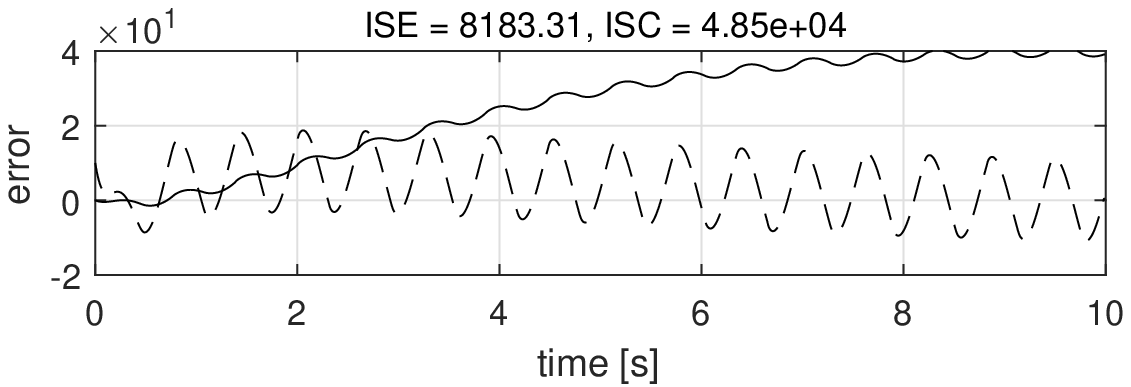}}\par
\subfloat[$\omega=1$]{\includegraphics[width=1\columnwidth]{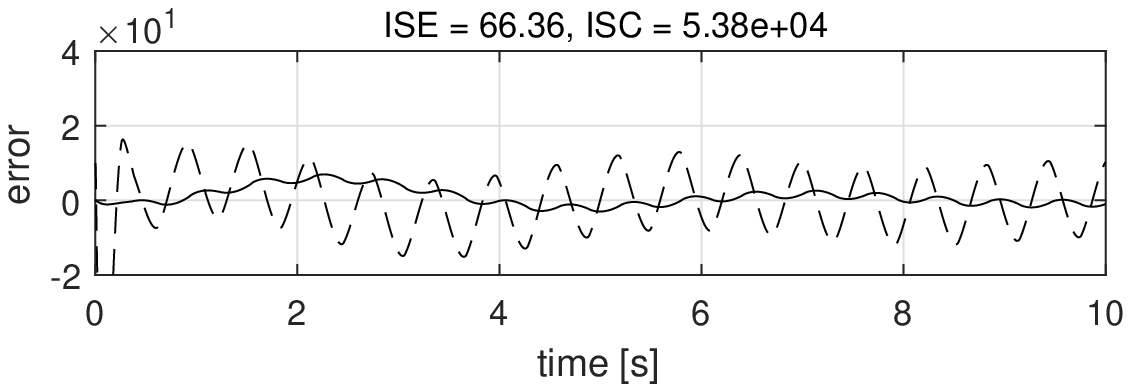}}\par
\subfloat[$\omega=4$]{\includegraphics[width=1\columnwidth]{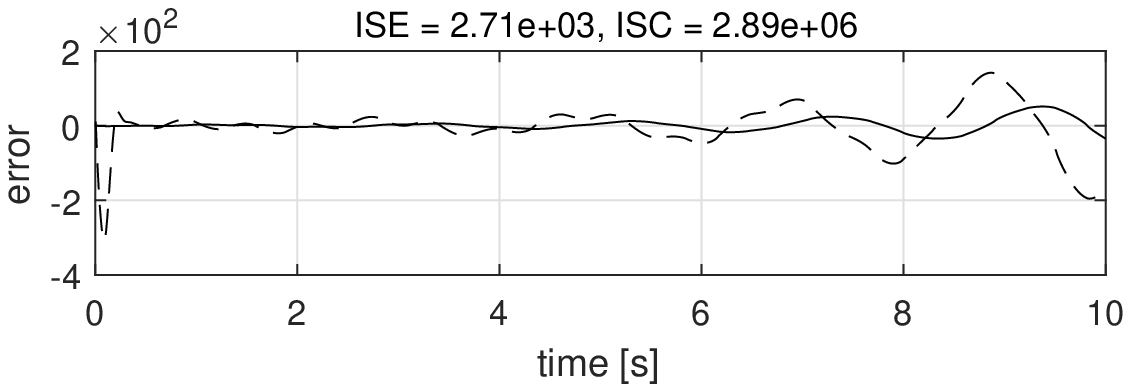}}\par
\caption{$T=\unit[1]{s}$, $\hat{z}_{3}$ disabled\label{fig:simulation:T1 pd}}%
\end{minipage}
\end{figure}

Selected results of simulations are presented in Figs.~\ref{fig:simulation:T01 adrc}-\ref{fig:simulation:T1 pd}. Tracking errors of two state variables
are presented on the plots. Error of $x_{1}$ is presented with solid
line, while $e_{2}$ has been plotted with dashed lines on each figure. 
Integrals of squared errors $e_1$ (ISE criterion) and integral of squared
control signals $v$ (ISC criterion) have been calculated for each simulation and are presented above the plots to quantify obtained tracking results. Tests were performed for different values of $T$ and $\omega$, as well as for compensation term $w_c=\hat{z}_{3}$ enabled or disabled, cf.  \eqref{eq:general:control law}. It can be clearly seen that the existence of some upper bound of $\Omega$ is confirmed by simulation results as proposed by Eq.  \eqref{eq:general:stability:Y1}.
As expected, value of this bound decreases with increase of time
constant $T$. In the conducted simulations it was not possible
to observe and confirm existence of any lower bound imposed on $\Omega$ and for an arbitrarily small $\omega$ stability of the system was being
maintained. Secondly, an influence of disturbance rejection term $\hat{z}_{3}$ is clearly visible and is twofold. For $\omega$ chosen to satisfy stability condition \ref{enum:general:stability:condition 1}, it can be observed, that the presence of the disturbance estimate allows significant decreasing of tracking errors $e_2$ caused by the input dynamics which were not modelled during the controller synthesis. Basically, a residual value of error $e_2$ becomes smaller for a higher value of bandwidth $\omega$. Error trajectory $e_1$ is also slightly modified, however, this effect is irrelevant according to ISE criterion. Nonetheless, usage
of the disturbance estimate leads to a significant shrink of $\Omega$ subset. It is plainly visible, that removal
of $\hat{z}_{3}$ estimate may lead to recovering of stability of the
system in comparison with simulation scenarios obtained using the corresponding ADRC
controller.

\section{Experimental results}

Practical experiments have been undertaken in order to further investigate
the considered control problem. All experiment were carried out using robotic
telescope mount developed at Institute of Automatic Control and Robotic
of Poznan University of Technology, \cite{KPKJPKBJN:2019}. The plant consists of a robotic
mount and an astronomic telescope with a mirror of diameter 0.5 m. The robotic
mount alone includes two axes driven independently by $\unit[24]{V}$
permanent magnet synchronous motors (PMSM) with high-precision ring
encoders producing absolute position measurement with 32-bit resolution.
Control algorithms has been implemented in C++ using Texas Instruments
AM4379 Sitara processor with ARM Cortex-A9 core clocked at $\unit[600]{MHz}$.
Beside control structure, prepared firmware contains several additional
blocks necessary for conducting of proper astronomical research.
Controller itself is implemented in a  cascade form which consists of independent
current and position loops. Both loops work simultaneously with frequency
of $\unit[10]{kHz}$. The current loop designed to precisely track desired
torque of the motor employs Park-Clark transformation of measured
phase currents to express motor dynamics in \emph{q-d }coordinated.
Both \emph{q }and \emph{d }axes are then controlled by independent
PI regulators with feedforward term and anti-windup correction which
satisfy the following equation
\begin{equation}
\begin{aligned}\dot{v} & =k_{i}\left(\tilde{i}-k_{s}\left(k_{p}\tilde{i}+v+u_{r}-\mathrm{sat}\left(k_{p}\tilde{i}+v+u_{r}\right)\right)\right),\\
u & =\mathrm{sat}\left(k_{p}\tilde{i}+v+u_{r},U_{m}\right),
\end{aligned}
\label{eq:experiments:current loop}
\end{equation}
where $\tilde{i}$ stands for current tracking error, $v$ is integrator
input signal, $u$ is regulator input, $u_{r}$ expresses feedforward
term, $k_{p}$, $k_{i}$ and $k_{s}$ are positive regulator gains,
and finally $\mathrm{sat}(u^{*},U_{m})$ is saturation function of
signal $u^{*}$ up to value of $U_{m}$. Output voltage $u$ is generated
using PWM output. Current in $d$ axis is stabilised at
 zero, while current of \emph{q} axis tracks the  desired
current of the axis. Relation between desired torque and desired current
is modelled as a constant gain equals $\unit[2.45]{\frac{Nm}{A}}$. Desired
torque is computed in the position loop by the active disturbance rejection
based controller designed for the second order mechanical system  modelled as follows
\begin{equation}
\left\{ \begin{array}{cl}
\dot{x}_{1} & =x_{2}\\
\dot{x}_{2} & =B\tau+\smash{\underbrace{f_{c}\cdot\mathrm{tanh}(f_{t}\cdot x_{2})}_{h(x_{2})}},
\end{array}\right.\label{eq:experiment:model}\\*[0.625\normalbaselineskip]
\end{equation}
where $x_{1}\in\mathbb{R}^2$ and $x_{2}\in\mathbb{R}^2$ are position and velocities of axes, $B$
is input matrix with diagonal coefficients equal $B_{1,1}=\frac{1}{5},B_{2,2}=\frac{1}{30}$,
$f_c$ is the constant positive Coulomb friction coefficient while $f_{t}=10^{3}$
expresses scaling term which defines steepness of friction model.
Velocity of the axis is approximated in the experiments
using either observer estimate $\hat{z}_{2}$ or desired trajectory
derivative $\dot{x}_{d}$. The assumed model of the friction force is strongly local, in the sense that different values of $f_{c}$
are required for different accelerations in a time instant when the sign of velocity
changes. This locality was overcame during the experiments by manual
changes of $f_{c}$ coefficient. While torque generated by the motor
is treated as an input signal of the mechanical system, there exists residual dynamics defined by the current loop which is not modelled in the position
loop. Here, we assume that this dynamics can be approximated by
\eqref{eq:general:input dynamics} and thus we can infer about the stability 
according to mathematical analysis considered in Section 2. Other disturbances come chiefly
from flexibility of the mount, ignored cross-coupling reactions between
joints and torque ripples generated by synchronous motors. Though some of these disturbances globally do not satisfy assumptions accepted for theoretical analysis of the system stability, in the considered scenario an influence of these dynamics is insignificant. Due to small desired velocities chosen in the experiment, these assumptions can be approximately satisfied here. All gains of the controllers chosen for experiments are collected in Table
\ref{tab:experiment:gains}.

\begin{table}[h]
\begin{centering}
\begin{tabular}{|c|c|c|}
\cline{2-3} \cline{3-3} 
\multicolumn{1}{c|}{} & Horizontal axis & Vertical axis\tabularnewline
\hline 
$K_{1}$ & $1.2\cdot10^{3}$ & $2.4\cdot10^{2}$\tabularnewline
\hline 
$K_{2}$ & $5.7\cdot10^{5}$ & $2.28\cdot10^{4}$\tabularnewline
\hline 
$K_{3}$ & $10^{8}$ & $0.8\cdot10^{6}$\tabularnewline
\hline 
$K_{p}$ & $225$ & $225$\tabularnewline
\hline 
$K_{d}$ & $24$ & $24$\tabularnewline
\hline 
\end{tabular}
\par\end{centering}
\caption{\label{tab:experiment:gains}Gains of the controllers and observers}
\end{table}

Here we present selected results of the experiments. In the investigated experimental scenarios both axes were at move simultaneously and the desired trajectory
was designed as a sine wave with period of $\unit[30]{s}$ and maximum
velocity of $50v_{s}$, in the first experiment, and $500v_{s}$, in the second, where $v_{s}=\unit[7.268\cdot10^{-5}]{\frac{rad}{s}}$ stands for the nominal velocity of stars on the night sky. 

During the system operation significant changes of friction forces are clearly visible and the influence of compensation term can be easily noticed. Since friction terms vary significantly around zero velocity the tracking accuracy is decreased. In such a case the process of the disturbance estimation is not performed fast enough. Furthermore, in the considered application one cannot select larger gains of the observer due to additional dynamics imposed by an actuator and delays in the control loop. Here, one can recall relationship \eqref{eq:control:conclusion} which clearly states that the tracking precision is dependent on the bound of $\Gamma_V$, cf. \eqref{eq:general:stability:Lyapunov perturbation}. Thus, one can expect that the tracking accuracy  increases in operating conditions when disturbances become slow-time varying. This is well illustrated in experiments where friction terms change in a wide range.

Each experiment presents results obtained with different approaches to $h_{u}$ term design. Once again integral squared error was calculated for each of 
the presented plots to ease evaluation of the obtained results.
\begin{figure}[h]
\centering
\begin{minipage}[t]{0.45\columnwidth}%
\subfloat[$h_{u}=0$ (no compensation)]{\includegraphics[width=1\columnwidth]{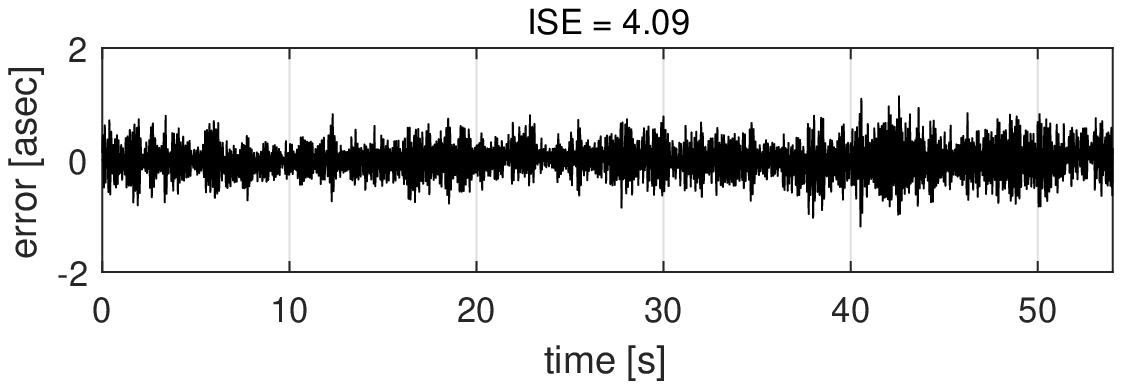}}\par
\subfloat[$h_{u}=0.5\cdot\tanh(\dot{x}_{d}\cdot10^{3})$]{\includegraphics[width=1\columnwidth]{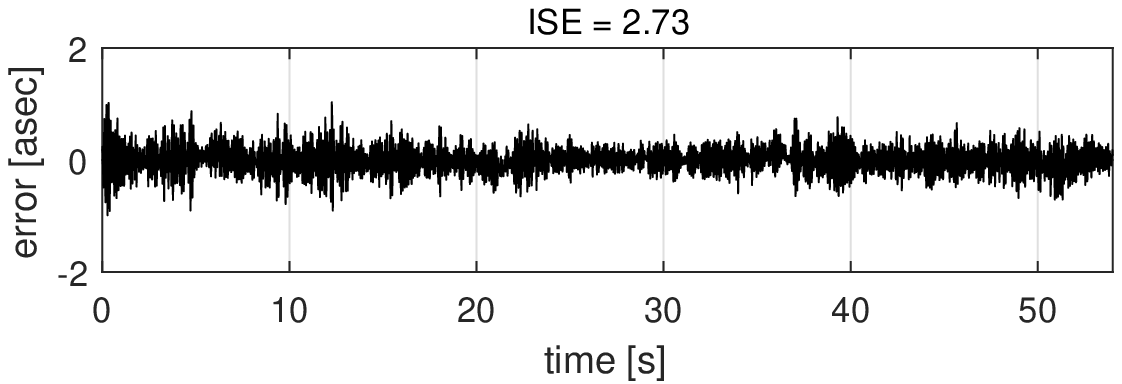}}\par
\subfloat[$h_{u}=0.5\cdot\tanh(\hat{z}_{2}\cdot10^{3})$]{\includegraphics[width=1\columnwidth]{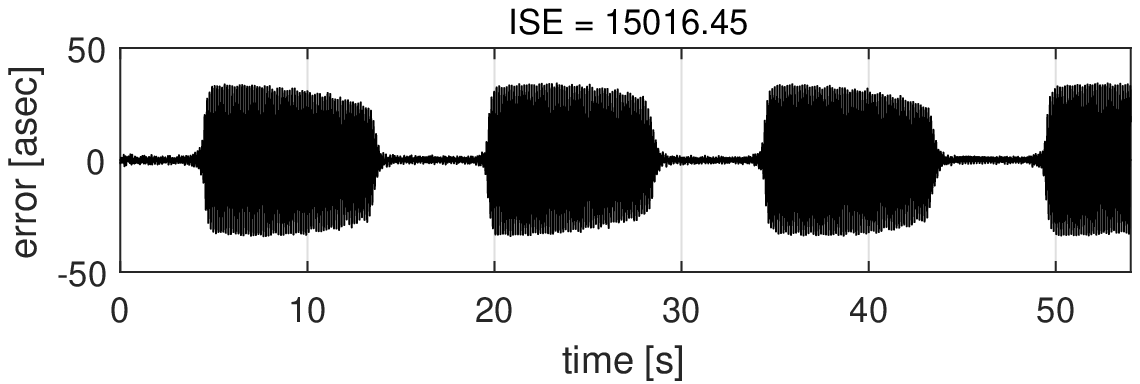}}\par
\caption{Horizontal axis, first experiment\label{fig:experiment:horizontal first}}
\end{minipage}%
\begin{minipage}[t]{0.45\columnwidth}%
\subfloat[$h_{u}=0$ (no compensation)]{\includegraphics[width=1\columnwidth]{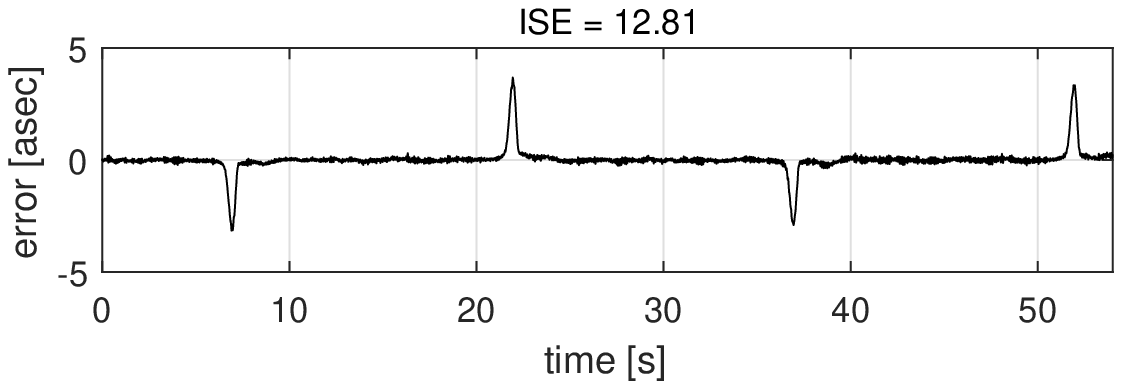}}\par
\subfloat[$h_{u}=0.5\cdot\tanh(\dot{x}_{d}\cdot10^{3})$]{\includegraphics[width=1\columnwidth]{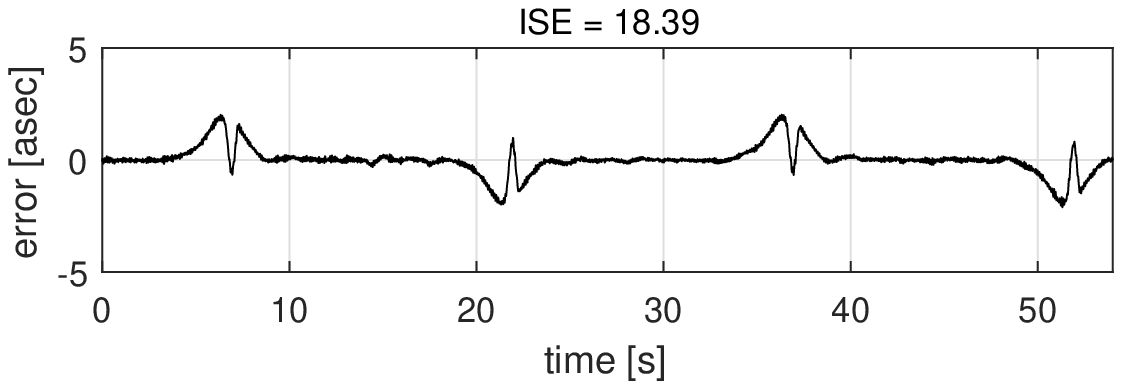}}\par
\subfloat[$h_{u}=0.5\cdot\tanh(\hat{z}_{2}\cdot10^{3})$]{\includegraphics[width=1\columnwidth]{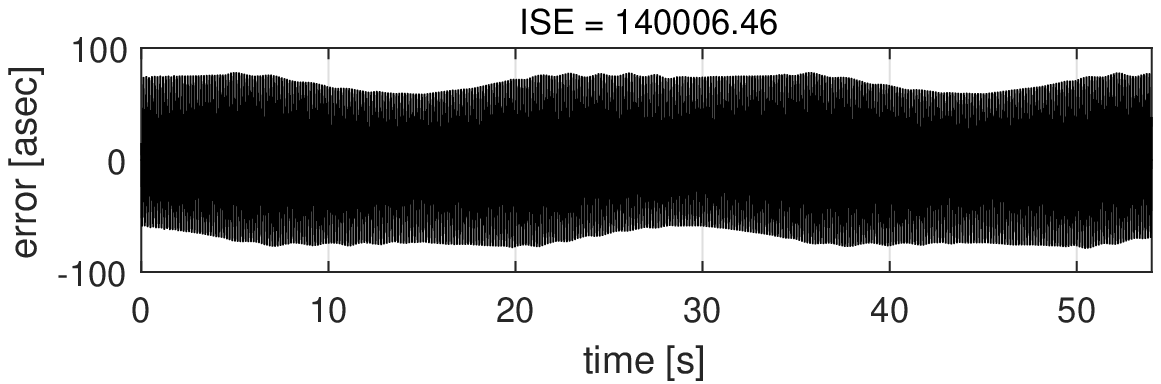}}\par
\caption{Vertical axis, first experiment\label{fig:experiment:vertical first}}
\end{minipage}
\end{figure}
\begin{figure}[h]
\centering
\begin{minipage}[t]{0.45\columnwidth}%
\subfloat[$h_{u}=0$ (no compensation)]{\includegraphics[width=1\columnwidth]{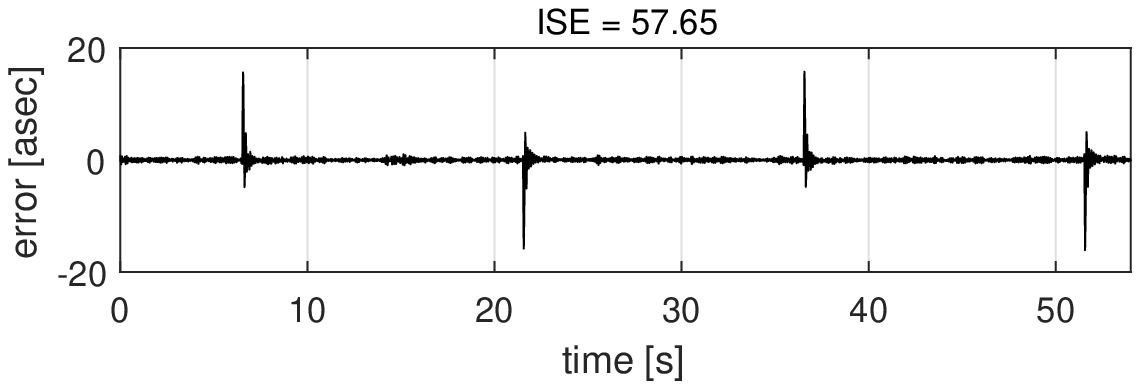}}\par
\subfloat[$h_{u}=0.15\cdot\tanh(\dot{x}_{d}\cdot10^{3})$]{\includegraphics[width=1\columnwidth]{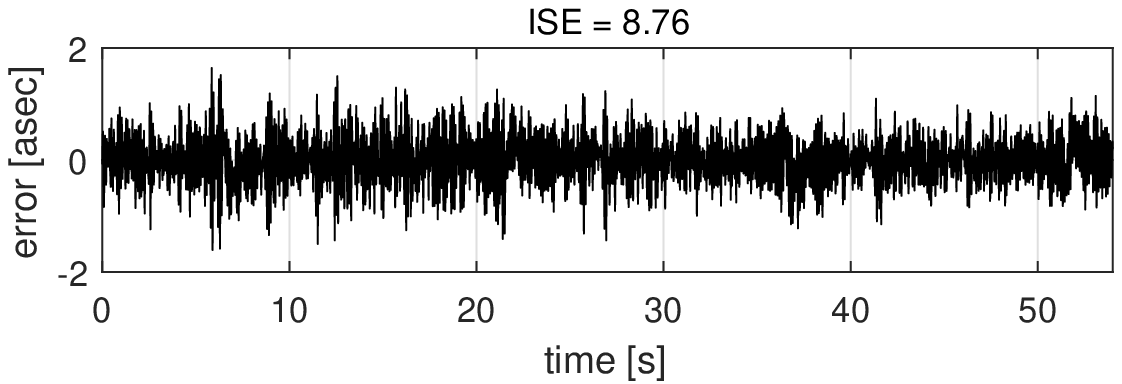}}\par
\caption{Horizontal axis, second experiment\label{fig:experiment:horizontal first-1}}
\end{minipage}%
\begin{minipage}[t]{0.45\columnwidth}%
\subfloat[$h_{u}=0$ (no compensation)]{\includegraphics[width=1\columnwidth]{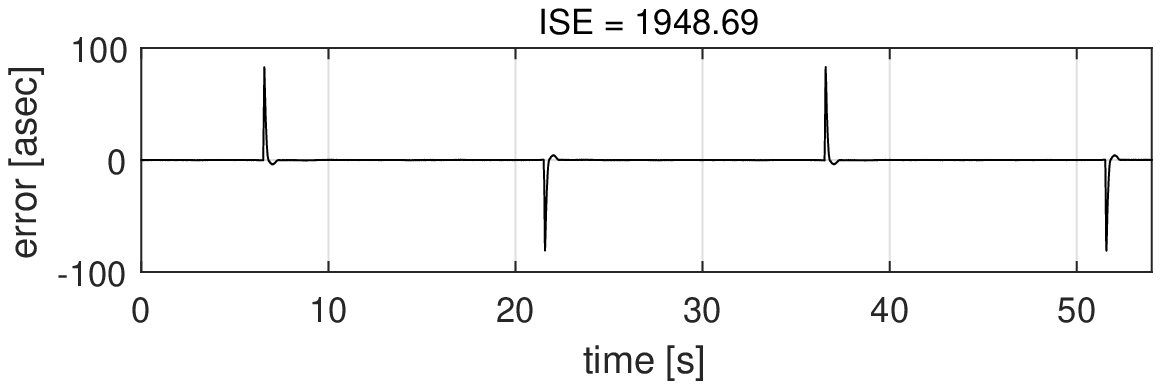}}\par
\subfloat[$h_{u}=0.15\cdot\tanh(\dot{x}_{d}\cdot10^{3})$]{\includegraphics[width=1\columnwidth]{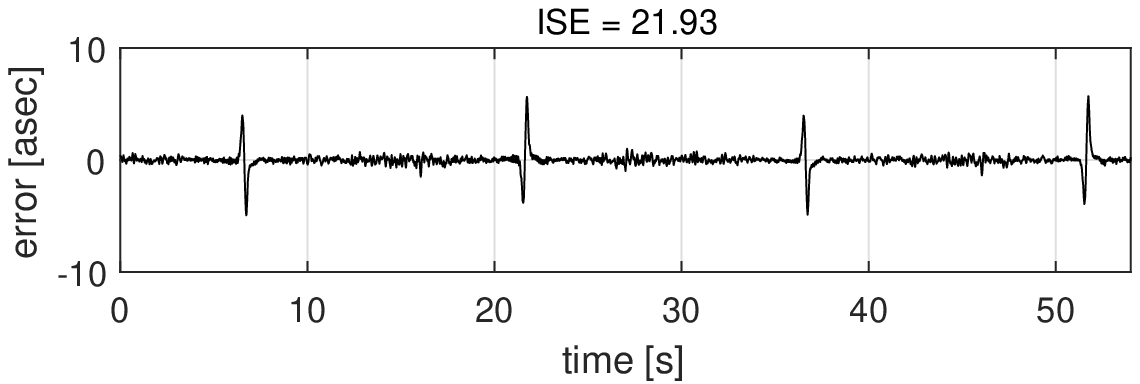}}\par
\caption{Vertical axis, second experiment\label{fig:experiment:vertical first-1}}
\end{minipage}
\end{figure}

Series of conclusions can be drawn from the presented results. Due to
inherently more disturbed dynamics of horizontal axis, any improvement
using friction compensation for slow trajectories is hardly achieved. Meanwhile,
the compensation term based on the desired trajectory, effectively decreases tracking error bound for all other experiments. As may be expected, 
compensation function based on estimates of state variables is unable
to provide any acceptable tracking quality due to inherent noise
in the signal and the existence of input dynamics.
It can be noted, that in the first experiment the friction compensation term allows one to decrease the bound of tracking error while overall quality expressed by ISE criterion
is worse in comparison to this obtained in experiment without the corresponding term in the feedback.
This behaviour is not seen in the second experiment, in which significant improvement 
was obtained for both axes in terms of error boundary as well as ISE criterion. 

\section{Conclusions}
This paper is focused on  the application of ADRC controller to a class of second order systems
subject to differentiable disturbances. In particular, the system is analysed taking into account the presence of the first order input dynamics and unmodelled terms which may include cross-coupling effects between the state variables. By the means of 
Lyapunov analysis, general conditions of practical stability are discussed. 
It is proved that, even in the presence of additional input dynamics, 
boundedness of partial derivatives of total disturbance can be a sufficient requirement to guarantee stability of the closed-loop system.

Using numerical simulations the considered controller is compared against a simple PD-based regulator. The obtained results confirm that in the case of input dynamics, the bandwidth of an extended observer is limited which restricts the effectiveness of the ADRC approach. Lastly, practical results of employing ADRC regulator in the task of trajectory tracking for  a robotised astronomical telescope mount are presented. In this application, it is assumed that friction effects are modelled inaccurately and a local drive control-loop is treated as unknown input dynamics. The obtained results illustrate that the considered control algorithm can provide a high tracking accuracy. 

Further research in this topic may include attempts to explore in more details conditions for the feasible selection of the observer parameters in order to guarantee the stability of the closed-loop system.
Other forms of input dynamics and observer models can also be considered in the future works.

\appendix
\section{Appendix}
\subsection{Selected properties of scaled dynamics}
Assuming that errors and gains are scaled according to \eqref{eq:general:regulator auxiliary errors}, \eqref{eq:general:observer auxiliary errors} and \eqref{eq:design:scaled_gains}
the following relationships are satisfied:
\begin{equation}
\begin{aligned}
\Delta_2\left(\kappa\omega\right)H_c \Delta_2^{-1}\left(\kappa\omega\right) =\kappa\omega\bar{H}_c,\ \Delta_3\left(\omega\right)H_o \Delta_3^{-1}\left(\omega\right) =\omega\bar{H}_o,\\ \Delta_2\left(\kappa\omega\right)W_1=W_1,\,
W_1\Delta_3^{-1}\left(\omega\right)=\bar{W}_1\Delta_3\left(\kappa\right)\\
W_2=\Delta_3\left(\kappa\omega\right)\bar{W}_2.\label{eq:app:scalled_terms}
\end{aligned}  
\end{equation}
\subsection{Computation of $\dot{v}$}
Taking advantage of estimate $\bar{z}$ and assuming that $w_c:=\hat{z}_3$ one can rewrite \eqref{eq:general:control law} as follows
\begin{equation}
v=B^{-1}\left(K_c e + K_c \begin{bmatrix}\tilde{z}_1^T&\tilde{z}_2^T \end{bmatrix}^T-h_u+\ddot{x}_d-\hat{z}_3\right),    
\end{equation}
where $K_c := \left[K_p\ K_d\right]$. Equivalently, one has
\begin{equation}
v=B^{-1}\left(K_c e - W_2\tilde{z}-h_u+\ddot{x}_d-z_3\right).    
\end{equation}
Consequently, time derivative of $v$ satisfies
\begin{equation}
\begin{aligned}
\dot{v}&=B^{-1}\left(K_c \dot{e}+W_2\dot{\tilde{z}}-\dot{h}_u+\dddot{x}_d-\dot{z}_3\right)\eqTwo{eq:general:regulator error dynamics}{eq:general:observator error dynamics}B^{-1}\left(K_c H_ce+K_cW_1\tilde{z}+K_c C_2B\tilde{u}+W_2H_o\tilde{z}\right.\\ &\quad\left.+W_2C_oB\tilde{u}+W_2C_1\dot{z}_3-\dot{z}_3-\dot{h}_u+\dddot{x}_d\right)=B^{-1}\left(K_c H_ce+K_cW_1\tilde{z}+W_2H_o\tilde{z}-\dot{h}_u+\dddot{x}_d\right).
\end{aligned}
\end{equation}

\subsection{Computations of $Y_3$ and $Y_4$}
By chain rule it can be shown that 
\begin{equation}
\dot{h}_1(z_1, z_2) = \begin{bmatrix}\frac{\partial h_{1}}{\partial z_{1}} & \frac{\partial h_{1}}{\partial z_{2}}\end{bmatrix}\begin{bmatrix}\dot{x}_{d}\\
\ddot{x}_{d}
\end{bmatrix}-\begin{bmatrix}\frac{\partial h_{1}}{\partial z_{1}} & \kappa\omega\frac{\partial h_{1}}{\partial z_{2}}\end{bmatrix}\dot{\bar{e}},\ 
\dot{h}_2(z_1,z_2) = \begin{bmatrix}\frac{\partial h_{2}}{\partial z_{1}} & \frac{\partial h_{2}}{\partial z_{2}}\end{bmatrix}\begin{bmatrix}\dot{x}_{d}\\
\ddot{x}_{d}
\end{bmatrix}-\begin{bmatrix}\frac{\partial h_{2}}{\partial z_{1}} & \kappa\omega\frac{\partial h_{2}}{\partial z_{2}}\end{bmatrix}\dot{\bar{e}},
\end{equation}
\begin{equation}
\dot{h}_1(\hat{z}_1,\hat{z}_2)=\begin{bmatrix}\frac{\partial h_{1}}{\partial\hat{z}_{1}} & \frac{\partial h_{1}}{\partial\hat{z}_{2}}\end{bmatrix}\begin{bmatrix}\dot{x}_{d}\\
\ddot{x}_{d}
\end{bmatrix}-\begin{bmatrix}\frac{\partial h_{1}}{\partial\hat{z}_{1}} & \kappa\omega\frac{\partial h_{1}}{\partial\hat{z}_{2}}\end{bmatrix}\dot{\bar{e}}-\begin{bmatrix}\frac{\partial h_{1}}{\partial\hat{z}_{1}} & \omega\frac{\partial h_{1}}{\partial\hat{z}_{2}} & 0\end{bmatrix}\dot{\bar{z}},\
\dot{h}_2(x_d,\dot{x}_d)=\begin{bmatrix}\frac{\partial h_{2}}{\partial x_{d}} & \frac{\partial h_{2}}{\partial\dot{x}_{d}}\end{bmatrix}\begin{bmatrix}\dot{x}_{d}\\
\ddot{x}_{d}
\end{bmatrix}.
\end{equation}
From here, following are true
\begin{equation}
\begin{aligned}
    \dot{h} - \dot{h}_u &= W_{h1}\begin{bmatrix}\dot{x}_{d}\\
\ddot{x}_{d}
\end{bmatrix}-W_{h2}\dot{\bar{e}}+W_{h3}\dot{\bar{z}},\ \dot{h}_u = W_{h4}\begin{bmatrix}\dot{x}_{d}\\
\ddot{x}_{d}
\end{bmatrix}-W_{h5}\dot{\bar{e}}-W_{h6}\dot{\bar{z}},
\end{aligned}
\end{equation}
what leads to solution of $Y_3$ by means of basic substitution.

Now, the computation of term $Y_4$ will be taken into account. Disturbance term $q(z_{1},z_{2},u,t)$ can be expressed in form of
\begin{equation}
\begin{aligned}
    \dot{q}(z_{1},z_{2},u,t)&=\frac{\partial q}{\partial z_{1}}\left(\dot{x}_{d}-\dot{e}_{1}\right)+\frac{\partial q}{\partial z_{2}}\left(\ddot{x}_{d}-\dot{e}_{2}\right)+\frac{\partial q}{\partial u}T^{-1}\left(-u+v\right)+\frac{\partial q}{\partial t}\\
&=W_{q1}\begin{bmatrix}\dot{x}_{d}\\
\ddot{x}_{d}
\end{bmatrix}+W_{q2}\dot{\bar{e}}-\frac{\partial q}{\partial u}T^{-1}\tilde{u}+\frac{\partial q}{\partial t}\\
&=W_{q1}\begin{bmatrix}\dot{x}_{d}\\
\ddot{x}_{d}
\end{bmatrix}+W_{q2}\left(\kappa\omega\bar{H}_{c}\bar{e}+\kappa^{-1}\omega\bar{W}_{1}D(\kappa)\bar{z}+\left(\kappa\omega\right)^{-1}C_{2}B\tilde{u}\right)-\frac{\partial q}{\partial u}T^{-1}\tilde{u}+\frac{\partial q}{\partial t}.
\end{aligned}
\end{equation}
\bibliographystyle{plain}
\bibliography{bibDP}

\begin{thebibliography}{10}

\bibitem{ACSA:2017}
C.~Aguilar-Ibanez, H.~Sira-Ramirez, and J.A. Acosta.
\newblock Stability of active disturbance rejection control for uncertain
  systems: {A} {L}yapunov perspective.
\newblock {\em International Journal of Robust and Nonlinear Control},
  27(18):4541--4553, 2017.

\bibitem{Bartol:1998}
G.~Bartolini, A.~Ferrara, and E.~Usai.
\newblock Chattering avoidance by second-order sliding mode control.
\newblock {\em IEEE Transactions on Automatic Control}, 43(2):241--246, Feb
  1998.

\bibitem{Bartol:2008}
G.~Bartolini, L.~Fridman, A.~Pisano, and E.~Usai, editors.
\newblock {\em Modern sliding mode control theory. New perspectives and
  applications}, volume~37 of {\em LNCIS}.
\newblock Springer-Verlag, 2008.

\bibitem{Bart:96}
A.~Bartoszewicz.
\newblock Time-varying sliding modes for second-order systems.
\newblock {\em IEE Proc. on Control Theory and Applications}, 143:455--462,
  1996.

\bibitem{Cast:2016}
I.~Castillo, L.~Fridman, and J.~A. Moreno.
\newblock Super-twisting algorithm in presence of time and state dependent
  perturbations.
\newblock {\em International Journal of Control}, 91(11):2535--2548, 2018.

\bibitem{CZG:2007}
Z.~Chen, Q.~Zheng, and Z.~Gao.
\newblock Active disturbance rejection control of chemical processes.
\newblock In {\em 2007 IEEE International Conference on Control Applications},
  pages 855--861, Oct 2007.

\bibitem{Fliess:2009}
M.~Fliess and C.~Join.
\newblock Model-free control and intelligent {PID} controllers: Towards a
  possible trivialization of nonlinear control?
\newblock {\em IFAC Proceedings Volumes}, 42(10):1531--1550, 2009.
\newblock 15th IFAC Symposium on System Identification.

\bibitem{FlJ:2013}
M.~Fliess and C.~Join.
\newblock Model-free control.
\newblock {\em {International Journal of Control}}, 86(12):2228--2252, December
  2013.

\bibitem{Gal:2015}
M.~Galicki.
\newblock Finite-time control of robotic manipulators.
\newblock {\em Automatica}, 51:49 -- 54, 2015.

\bibitem{Gal:2016}
M.~Galicki.
\newblock Finite-time trajectory tracking control in a task space of robotic
  manipulators.
\newblock {\em Automatica}, 67:165--170, 2016.

\bibitem{Gao:2002}
Z.~Gao.
\newblock From linear to nonlinear control means: A practical progression.
\newblock {\em ISA Transactions}, 41(2):177 -- 189, 2002.

\bibitem{Gao:2006}
Z.~Gao.
\newblock Active disturbance rejection control: a paradigm shift in feedback
  control system design.
\newblock In {\em 2006 American Control Conference}, pages 2399--2405, June
  2006.

\bibitem{Han:1998}
J.~Han.
\newblock Auto-disturbance rejection control and its applications.
\newblock {\em Control and Decision}, 13(1), 1998.
\newblock In Chinese.

\bibitem{Han:2009}
J.~Han.
\newblock From {PID} to {A}ctive {D}isturbance {R}ejection {C}ontrol.
\newblock {\em IEEE Transactions on Industrial Electronics}, 56(3):900--906,
  March 2009.

\bibitem{KhP:2014}
H.K. Khalil and L.~Praly.
\newblock High-gain observers in nonlinear feedback control.
\newblock {\em International Journal of Robust and Nonlinear Control},
  24(6):993--1015, 2014.

\bibitem{KPKJPKBJN:2019}
K.~Koz{\l}owski, D.~Pazderski, B.~Krysiak, T.~Jedwabny, J.~Piasek,
  S.~Koz{\l}owski, S.~Brock, D.~Janiszewski, and K.~Nowopolski.
\newblock High precision automated astronomical mount.
\newblock In R.~Szewczyk, C.~Zieli\'nski, and M.~Kaliczy\'nska, editors, {\em
  Automation 2019. Advances in Intelligent Systems and Computing}, volume 920.
  Springer, 2020.

\bibitem{Levant:1993}
A.~Levant.
\newblock Sliding order and sliding accuracy in sliding mode control.
\newblock {\em International Journal of Control}, 58(6):1247--1263, 1993.

\bibitem{Levant:1998}
A.~Levant.
\newblock Robust exact differentiation via sliding mode technique.
\newblock {\em Automatica}, 34(3):379--384, March 1998.

\bibitem{MiH:2015}
R.~Mado\'nski and P.~Herman.
\newblock Survey on methods of increasing the efficiency of extended state
  disturbance observers.
\newblock {\em ISA Transactions}, 56:18--27, 2015.

\bibitem{MiGao:2005}
R.~Miklosovic and Z.~Gao.
\newblock A dynamic decoupling method for controlling high performance turbofan
  engines.
\newblock {\em IFAC Proceedings Volumes}, 38(1):532--537, 2005.
\newblock 16th IFAC World Congress.

\bibitem{NVMPB:2012}
A.~Nowacka-Leverton, M.~Micha{\l}ek, D.~Pazderski, and A.~Bartoszewicz.
\newblock Experimental verification of smc with moving switching lines applied
  to hoisting crane vertical motion control.
\newblock {\em ISA transactions}, 51:682--693, 2012.

\bibitem{NSKCFL:2018}
P.~Nowak, K~Stebel, T.~K{\l}opot, J.~Czeczot, M.~Fr{\k{a}}tczak, and
  P.~Laszczyk.
\newblock Flexible function block for industrial applications of active
  disturbance rejection controller.
\newblock {\em Archives of Control Sciences}, 28(3):379--400, 2018.

\bibitem{SiGao:2017}
S.~Shao and Z.~Gao.
\newblock On the conditions of exponential stability in active disturbance
  rejection control based on singular perturbation analysis.
\newblock {\em International Journal of Control}, 90(10):2085--2097, 2017.

\bibitem{SiGao:2005}
B.~Sun and Z.~Gao.
\newblock A dsp-based active disturbance rejection control design for a 1-kw
  h-bridge dc-dc power converter.
\newblock {\em IEEE Transactions on Industrial Electronics}, 52(5):1271--1277,
  Oct 2005.

\bibitem{Utk:77}
V.~Utkin.
\newblock Variable structure systems with sliding modes.
\newblock {\em IEEE Trans. on Automatic Control}, 22:212--222, 1977.

\bibitem{WCW:2007}
D.~Wu, K.~Chen, and X.~Wang.
\newblock Tracking control and active disturbance rejection with application to
  noncircular machining.
\newblock {\em International Journal of Machine Tools and Manufacture},
  47(15):2207--2217, 2007.

\end{thebibliography}
\end{document}